\documentclass[preprint,12pt]{aastex}
\def\um{$\mu$m}
\begin{document}
\title{Spitzer 24 \um\ Survey for Dust Disks around Hot White Dwarfs}
%
%
\author{You-Hua Chu\altaffilmark{1}, Kate Y.~L.\ Su\altaffilmark{2},
Jana Bilikova\altaffilmark{1}, Robert A.\ Gruendl\altaffilmark{1},
Orsola De Marco\altaffilmark{3}, Martin A.\ Guerrero\altaffilmark{4}, 
Adria C.\ Updike\altaffilmark{5,6,7}, Kevin Volk\altaffilmark{8}, 
and Thomas Rauch\altaffilmark{9}}
\altaffiltext{1}{\itshape Department of Astronomy, University of Illinois
at Urbana-Champaign, 1002 West Green Street, Urbana, IL 61801, USA;
yhchu@illinois.edu} 
\altaffiltext{2}{\itshape Stewart Observatory, University of Arizona,
Tucson, AZ 85721, USA} 
\altaffiltext{3}{\itshape Macquarie University, Sydney, NSW 2109, Australia}
\altaffiltext{4}{\itshape Instituto de Astrof\'{\i}sica de Andaluc\'{\i}a,
CSIC. c/ Camino Bajo de Hu\'etor 50, E-18008 Granada, Spain}
\altaffiltext{5}{\itshape CRESST and the Observational Cosmology Laboratory, 
NASA/GSFC, Greenbelt, MD 20771, USA}
\altaffiltext{6}{\itshape Department of Astronomy, University of Maryland, 
College Park, MD 20742, USA}
\altaffiltext{7}{\itshape Department of Physics and Astronomy,
Clemson University, Clemson, SC 29634-0978, USA}
\altaffiltext{8}{\itshape Space Telescope Science Institute,
3700 San Martin Drive, Baltimore, MD 21218, USA} 
\altaffiltext{9}{\itshape Institut f\"ur Astronomie und Astrophysik
 T\"ubingen (IAAT), Abteilung Astronomie, Sand 1, D-72076 T\"ubingen,
Germany} 
%
%

%
%
%
\begin{abstract}
Two types of dust disks around white dwarfs (WDs) have been reported:
small dust disks around cool metal-rich WDs consisting of tidally disrupted
asteroids, and a large dust disk around the hot central WD of the
Helix planetary nebula (PN) possibly produced by collisions among 
Kuiper Belt-like objects.  
To search for more dust disks of the latter type, we have conducted 
a {\it Spitzer} MIPS 24 \um\ survey of 71 hot WDs or pre-WDs, among
which 35 are central stars of PNe (CSPNs).  
Nine of these evolved stars are detected and their 24 \um\ flux 
densities are at least two orders of magnitude higher than their
expected photospheric emission.  
Considering the bias against detection of distant objects, the
24 \um\ detection rate for the sample is $\gtrsim$15\%.
It is striking that seven, or $\sim$20\%, of the WD and pre-WDs in 
known PNe exhibit 24 \um\ excesses, while two, or 5--6\%, of the WDs 
not in PNe show 24 \um\ excesses and they have the lowest 24 \um\
flux densities.
We have obtained follow-up {\it Spitzer} IRS spectra for five objects.  
Four show clear continuum emission at 24 \um, and one is overwhelmed by
a bright neighboring star but still show a hint of continuum emission.
In the cases of WD\,0950+139 and CSPN K\,1-22, a late-type companion
is present, making it difficult to determine whether the excess 24 \um\ 
emission is associated with the WD or its red companion.
High-resolution images in the mid-IR are needed to establish
unambiguously the stars responsible for the 24 \um\ excesses.
\end{abstract}
\subjectheadings{infrared: stars -- circumstellar matter -- white dwarfs
 -- planetary nebulae: general}
\maketitle
%
%
%
%
\section{Introduction}  \label{sec:intro}

The {\it Spitzer Space Telescope} \citep{Wetal04}, with its superb
sensitivity and resolution at infrared (IR) wavelengths, provides an 
excellent opportunity to study planetary debris disks around stars
\citep{Suetal06,Tetal08,Caetal09}.  
For instance, a comprehensive {\it Spitzer} 24 \um\ survey of 
main-sequence A-type stars has shown that up to $\sim$50\% of young 
($\lesssim$30 Myr) stars have little or no 24 \um\ excess emission 
from debris disks, large debris-disk excesses decrease significantly 
at ages of $\sim$150 Myr, and much of the dust detected may be 
generated episodically by collisions of large planetesimals 
\citep{Rietal05}.  The dust in these debris disks would have 
dissipated long before the stars evolve off the main sequence.

Dust can be replenished during late evolutionary stages.
As a low- or intermediate-mass star loses a significant fraction 
of its initial mass to become a white dwarf (WD), its planetary 
system expands.  Sub-planetary objects, such as asteroids and comets,
can be injected to very small radii and be tidally pulverized by
the WD, while orbital resonances with giant planets can
raise the collision rates among sub-planetary objects and 
generate dust \citep{DS02}.  
This freshly produced dust can be detected through IR excesses and 
allows us to peer into the late evolution of planetary systems.
 
The first two WDs reported to possess dust disks were G29-38 and GD\,362,
both exhibiting near-IR excesses that were confirmed spectroscopically
to be dust continuum emission \citep{ZB87, Betal05, Ketal05,Reetal05}.
A subsequent {\it Spitzer} survey of 124 WDs at 4.5 and 8.0 $\mu$m
found one additional dust disk, around WD\,2115$-$560 \citep{Metal07,
vHetal07}.
As G29-38, GD\,362, and WD\,2115$-$560 are WDs with photospheric 
absorption lines of heavy elements, searches for dust disks have been 
conducted for DAZ and DBZ WDs, and indeed many more dust disks were 
discovered.
For example, a near-IR spectroscopic survey of 
20 DAZ WDs found a dust disk around GD\,56 \citep{Ketal06}, the
presence of a dust disk around the DAZ WD\,1150$-$153 was diagnosed by
$K$-band excess and confirmed spectroscopically \citep{KR07}, and 
{\it Spitzer} photometric observations of 9 DAZ/DBZ WDs revealed dust 
disks around GD\,40, GD\,133, and PG\,1015+161 \citep{Jetal07}.  
One common characteristic of these dust disks around
DAZ/DBZ WDs is that they are all small, with outer radii $\ll$0.01 AU.
As the dust disks are completely within the Roche limits of the WDs
and the dust mass is estimated to be only $\sim$10$^{18}$ g, 
it is suggested that tidally disrupted asteroids produce the dust 
disks and that the accretion of this dust enriches refractory metals,
such as Ca, Mg, Fe, and Ti, in the WD atmospheres \citep{Jura03,Jetal07,
Zetal07}.
To date, $\sim$20 dust disks around DAZ/DBZ WDs have been reported, all
consistent with this suggested origin of tidal disruption of asteroids
\citep{Fetal10}.  More dust disks of this type are being found 
from the {\it Wide-Field Infrared Survey Explorer} observations of WDs 
\citep[e.g.,][]{Detal11}.

An entirely different kind of dust disk has been discovered around
the central WD of the Helix planetary nebula (PN), WD\,2226$-$210
\citep{Suetal07}.
{\it Spitzer} observations of the Helix Nebula show a bright compact
source coincident with the central WD in the 24 and 70 \um\ bands, and
follow-up Infrared Spectrograph \citep[IRS;][]{Hetal04} observations 
have verified that the mid-IR emission originates from a dust continuum.  
The spectral energy distribution (SED) of this IR-emitter indicates a 
blackbody temperature of 90-130 K, and its luminosity, 
5-11$\times$10$^{31}$ ergs s$^{-1}$, 
requires an emitting area of 3.8-38 AU$^2$.  These properties can only
be explained by the presence of a dust cloud; furthermore, little 
extinction exists toward the WD, so the dust cloud must be flattened
with a disk geometry.
Adopting a stellar effective temperature of 110,000 K for 
WD\,2226$-$210 \citep{Nap99} and assuming astronomical silicates
with a power-law size distribution and a maximum grain radius
of 1000 \um\ for the dust grains, models of
the SED indicate that the dust disk extends between 35 and 150 AU 
from the WD and has a mass of $\sim$0.13 $M_\oplus$.
Since this dust must have been generated recently and since the
radial location of the dust disk corresponds to that of the Kuiper 
Belt in the Solar System, \citet{Suetal07} suggest that the dust 
disk around WD\,2226$-$210 was produced by collisions of Kuiper
Belt-like objects (KBOs) or the break-up of comets.

To simulate the dust disk of WD\,2226$-$210, the dynamic evolution
of a debris disk around a 3 $M_\odot$ star has been modeled from the
main sequence (corresponding to $\sim$A0\,V) to the WD stage, and 
it is found that collisions among KBOs may produce
the amount of dust observed \citep{BW10,Detal10}.
If the dusk disk around the Helix central star is indeed produced 
by collisions of KBOs, similar dust disks should be found around
other WDs and a survey would allow us to assess their frequency of 
occurrence and physical properties.  These results can then be
compared with models of debris disks evolution \citep{BW10,Detal10}
for implications on their planetary systems.

To search for dust disks similar to that around the central WD of
the Helix Nebula, we have conducted three surveys using {\it Spitzer}
observations:
(1) 24 \um\ survey of hot WDs and pre-WDs (this paper),
(2) archival survey of IR excesses of WDs \citep{Cetal11,Retal11},
and (3) archival survey of IR excesses of central stars
of PNe \citep[CSPNs;][]{Betal11a}. 
The combined results from these three surveys will provide a 
comprehensive picture of post-main sequence dust production and 
dynamic evolution of debris disks.  
This paper reports the results of the {\it Spitzer} 24
\um\ survey of hot WDs as well as follow-up spectroscopic and imaging 
observations for a subset of hot WDs with 24 \um\ excesses.
Section 2  describes the target selection and observations, Section 3
reports the results, and Section 4 discusses the implications.  A
summary is given in Section 5.

\section{Observations and Data Reduction}  \label{sec:obs}

The large dust disk around the central WD of the Helix Nebula 
is detectable at 24 \um\ because WD\,2226$-$210 has a high
temperature and thus high luminosity.
Stellar luminosity is a steeper function of temperature than of
radius, and WDs span a small range of radii; furthermore,
stellar effective temperatures are more readily available in
the literature than distances and luminosities.  Therefore, we
use stellar effective temperature as our main target selection
criterion and selected targets from two sources for our 24 
\um\ survey.
First, we use the web version of the \citet{MS99} WD Catalog 
to search for WDs whose spectral classifications indicate 
temperatures near or higher than 100,000 K; 58 such hot WDs
are found.
Hot WDs are the youngest WDs and often surrounded by evolved PNe;
22 of the 58 WDs selected are in known PNe.
To probe whether dust disks are present around pre-WDs,
we have selected 14 hot CSPNs \citep{Nap99} whose effective 
temperatures are $\ge$100,000 K but whose surface gravity 
is not yet high enough to be classified as WDs. 
Although pre-WDs evolve along tracks of almost constant 
luminosities in the HR diagram, we do not select CSPNs with 
lower effective temperatures as they are usually surrounded by
smaller and brighter PNe, making it difficult to obtain accurate
photometric measurements of the CSPNs.
These 58 WDs and 14 pre-WDs form the target list of our {\it Spitzer}
24 \um\ survey for dust disks.

Among the 72 targets, ``WD 0123$-$842'' was a misnomer from wrong 
coordinates, and has been removed from the web version of the 
McCook \& Sion WD Catalog (E.\ Sion, personal communication); thus,
this spurious object will not be discussed further in this paper.
The 71 valid targets are listed in Table 1.  The coordinates in 
columns 2-3 are measured from the Digitized Sky Survey 2 (DSS2),
and in many cases represent an improved accuracy when compared with
those given by McCook \& Sion or SIMBAD. 
Unless noted, the effective temperatures ($T_{\rm eff}$) in 
column 4 are taken from \citet{Nap99} or converted from the 
spectral classification in column 5 taken from McCook \& 
Sion's WD catalog.  
The common name and PNG number of associated PNe are given in columns 6-7.

The {\it Spitzer} observations reported in this paper were obtained 
from three programs.  
The main program is our 24 \um\ survey of the hot WDs or pre-WDs 
(Program 40953) using the Multiband Imaging Photometer 
for {\it Spitzer} \citep[MIPS;][]{Retal04}. 
To assess the nature of hot WDs' 24 \um\ excesses, we
have further obtained {\it Spitzer} IRS, Infrared Array Camera
\citep[IRAC;][]{fazio04}, and MIPS 70 \um\ observations 
(Program 50629).
As the MIPS 24 \um\ survey was not completed before the proposal 
deadline for {\it Spitzer} Cycle 5, the last cryogenic cycle of the
{\it Spitzer} mission, IRS and MIPS 70 \um\ observations were 
obtained for only a subset of hot WDs with 24 \um\ excesses:  
CSPN K\,1-22, WD\,0103+732 (CSPN EGB\,1), WD\,0127+581 (CSPN Sh\,2-188),
and WD\,0439+466 (CSPN Sh\,2-216).
Among these four targets, archival IRAC observations are available
for WD\,0439+466; we have thus obtained new IRAC observations for only
the other three.
Finally, we have obtained a MIPS 70 \um\ observation of 
KPD\,0005+5106 (= WD\,0005+511) through a {\it Chandra X-ray Observatory}
observing program (GO8-9026).  This MIPS observation was merged with 
our {\it Spitzer} Program 40953 for convenience.

\subsection{MIPS Observations}

\subsubsection{MIPS 24 $\mu$m Observations}
In the MIPS 24 \um\ survey of hot WDs, each of the 71 objects in Table 1 
was imaged using the small-field photometry mode which obtained a 
sequence of 14 dithered exposures in a preset pattern (see the {\it Spitzer
Space Telescope} Observer's Manual for more details).  
The observations used an exposure time of 10~s and cycled through the 
pattern 3 times, yielding a total exposure time of 420~s in the central
3\farcm2$\times$3\farcm2 region.
In a typical median background condition, the depth of the survey will
reach a 1-$\sigma$ point-source sensitivity of 33.6~$\mu$Jy.
The raw data were processed using the Data Analysis Tool 
\citep[DAT;][]{Getal05} for basic reduction (e.g., dark 
subtraction and flat fielding). 
All individual exposures were first corrected with a
scan-mirror-dependent flat to correct for the dark spots due to 
particulate matter on the pick-off mirror. For targets that have no 
large-scale extended
emission (from either the surrounding PNe or the background Galactic 
cirrus), a second flat field generated from the data itself was applied
to correct the possible dark latency and scattered
light gradient in order to enhance photometric sensitivity. Details of
these processing procedures can be found in \citet{Enetal07}. The final
mosaics were then constructed with a pixel size of 1\farcs245 (half the
size of the physical pixel) for photometry measurements. The
calibration factor 4.54$\times$10$^{-2}$ MJy sr$^{-1}$ (DN
s$^{-1}$)$^{-1}$ was used to convert the instrumental unit to 
physical units \citep{Enetal07}.

All final mosaics were astrometrically calibrated by comparisons with
the Two Micron All Sky Survey \citep[2MASS;][]{Setal06} sources in the
field to establish the World Coordinate System (WCS), the accuracy of
which is generally better than $\sim$1\arcsec.  We first performed
source extraction using {\it StarFinder} \citep{diolaiti00} with a
smoothed theoretical point spread function (PSF) generated by the
STinyTim program (Krist 2006). We consider the WD detected if
the extracted source position is within 1\farcs5 from the given WD
position.  In other words, this method ensures a point-source-like
object is required to be coincident with the given WD position.
For sources that are not detected, we estimate their
``observed'' point source flux by fixing the PSF at the given WD
position on the PSF-subtracted (source-free) image and using the
minimum $\chi^2$ technique. In some cases where the nebular emission
is bright, this PSF flux is totally dominated by the background (e.g.,
CSPN IC 289, CSPN MeWe 1-3, and WD\,1958+015 in NGC\,6852). 
These PSF fluxes (non-color corrected) are listed in column 3 of
Table 2.

We have estimated the 1-$\sigma$ point
source sensitivity of the observation using the pixel-to-pixel
variation inside a radius of 44\arcsec~centered at the source position
on the source-free image (i.e., the systematic background noise). The
final photometry error includes (1) the systematic error of the
observation, estimated from pixel-to-pixel variations; (2) 24 $\mu$m
detector repeatability, $\sim$1\% of the measured flux; and (3) 24 $\mu$m
confusion noise, $\sim$15 $\mu$Jy, a median value estimated from the
extragalactic source counts \citep{dole04}. These error contributions are
summed in quadrature and listed in column 4 of Table 2 as ``uncertainty''. 
This ``uncertainty'' has no direct relationship to the detectability of
the source. It simply reflects the photometric accuracy if a point
source was coincident with the given WD position.
For non-detections, we also list 3-$\sigma$ upper limits in column 5 of
Table 2, computed as the three times of the uncertainty plus the PSF flux.
In most cases, these 3-$\sigma$ upper limits are totally dominant by the 
surrounding bright nebular emission as remarked in column 8 in Table 2.

For detected sources, we have also carried out aperture photometry using
a source aperture of 6\farcs225 radius, a sky annulus of radii 
19\farcs92 -- 29\farcs88, and an aperture correction of 1.7.
These results are compared with those determined from the PSF-fitting method.
When the two measurements for a source are discrepant, we examine the field
for nearby objects or background structures that might compromise the
photometry, and adopt the measurement that is less compromised. 
The aperture photometry result is adopted for WD\,0439+466, and the 
PSF-fitting results are adopted for the other eight cases of detections.
When IRS observations are available, we use the background-subtracted 
flux densities near 24 \um\ to further constrain the MIPS 24 \um\
photometric measurements.

\subsubsection{MIPS 70 $\mu$m Observations}

Four targets were included in the MIPS 70 \um\ observations as
part of our Program 50629. 
The three original targets, CSPN K\,1-22, WD\,0103$+$732, and
WD\,0439$+$466, were observed in the photometry raster map 
mode with 3$\times$1 maps in the array column and row direction
and a step size of 1/8 array.
This yields a uniformly covered area within a diameter of 
3\farcm2 centered at the target. 
Each map was repeated 4 times (cycles) with 10 s integration
per frame, resulting in a total effective integration of 960 s.
WD\,0005+511 was a target from the {\it Chandra} program GO8-9026.
Its 70 \um\ observations were made in the default photometry mode,
with 10 s integration per frame and three cycles, and the
total integration time was 240 s.

The basic reduction (dark subtraction, illumination correction) of 
the 70 \um\ data was processed using DAT, similar to that of the
24 \um\ data.  The known transient behavior in the 70 \um\ array
was removed by masking out the sources in the field of view and time 
filtering the data \citep{Getal07}. For the objects in PNe
with extended nebulosity seen at 70 \um, the size of the masked 
area was adjusted to cover most of the nebulosity accordingly. 
The final mosaics were combined using the WCS with a subpixel size 
of 4\farcs93.  Figure 1 shows the 70 \um\ images alongside the 
24 \um\ images of K\,1-22, WD\,0103+732, and WD\,0439+466.

The 70 \um\ image of WD\,0103+732 shows bright extended emission with a 
central peak at the expected WD position. The diffuse emission at 70 \um,
appearing more extended than that in the 24 \um\ band, is most likely
dominated by bright emission lines such as the [\ion{O}{3}] 88 \um\ line.
To minimize the nebular contamination, we performed aperture photometry with
a very small (16\arcsec) aperture and sky annulus of 18\arcsec--39\arcsec.
We have also used PSF fitting to estimate the source brightness. 
Both methods give a point source flux of $\sim$55 mJy.  The pixel-to-pixel 
variations in the data suggest 1-$\sigma$ point source sensitivity of 8.4 mJy. 

WD\,0439+466 is coincident with a very faint 70 \um\ source.
The source is superposed on diffuse emission, similar to the case 
of WD\,0103+732.
Using the same approach as for WD\,0103+732, we estimate the source
to have a 70 \um\ flux density of 9.0$\pm$8.4 mJy.  

Unlike the above two WDs, CSPN K\,1-22 has no obvious source coincident
with the WD. The data were very noisy, probably due to many foreground
and background sources (evident at 24 \um\ as well). The estimated
1-$\sigma$ point source sensitivity is $\sim$4 mJy.  WD\,0005+511 is not
detected at 70 \um, either; its 1-$\sigma$ sensitivity is 5.4 mJy.

\subsection{IRAC observations}

IRAC observations of three targets, CSPN K\,1-22, WD\,0103$+$732, 
and WD\,0127$+$581 were obtained at 3.6, 4.5, 5.8 and 8.0~\um\ 
in our Program 50629.  These observations all used a 30~s frame time 
and a cyclic dither pattern with medium offsets to obtain 5 frames 
for a total integration time of 150~s for each target.  The basic 
calibrated data (BCD) frames from these observations were reduced 
using standard routines within the MOPEX package.

We have measured flux densities in each of the IRAC bands using
the IRAF task \textsf{phot} to perform aperture photometry.  
These measurements used a 3\farcs6 radius source aperture with a 
background estimated from the surrounding annulus of radii 3\farcs6
to 8\farcs4.  An aperture correction based on the results 
tabulated in the IRAC Data Handbook (ver 3.0) was then applied to 
obtain the flux densities reported in Table~3.

Archival IRAC observations are available for WD\,0439+466 (Program 
30432; PI: Burleigh), WD\,0726+133 (Program 30285; PI: Fazio), 
and the CSPN NGC\,2438 (Program 68; PI: Fazio).
These archival data were downloaded and reduced in the same manner as our
own observations.  Their flux densities are also reported in Table~3.

\subsection{IRS Observations}

We have also obtained follow-up IRS observations for CSPN K\,1-22, 
WD\,0103+732, WD\,0127+581, and WD\,0439+466 in our Program 50629.  
All sources were observed using the low-resolution modules 
SL1 (7.4--14.5 \um), SL2 (5.2--7.7 \um), LL1 (19.5--38.0 \um),
and LL2 (14.0--21.3 \um).  
See \citet{Hetal04} for a more detailed description of the IRS and its 
capabilities.  CSPN K\,1-22 and WD\,0103+732 were observed in the IRS 
staring mode after a peakup on a nearby source.  WD\,0127+581, the
faintest target, was observed in mapping mode where the target was 
sequentially stepped along the slit to facilitate an improved background 
subtraction.  The SL1 and SL2 observations of WD\,0127+581 used 8 pointings 
spaced by 3\arcsec\ while the LL1 and LL2 observations used 12 
pointings spaced by 6\arcsec.  The MIPS 24~\um\ observations of
WD\,0439+466 showed diffuse emission around the WD,
thus the IRS observations of WD\,0439+466 were made in the mapping mode 
to obtain spectra of the WD and the diffuse emission by taking
spectra from a series of dense slit positions centered on the WD 
and sequentially offset in the direction perpendicular to the slit.
Specifically, the SL1 and SL2 maps were comprised of 9 pointings spaced by 
3\farcs6 while the LL1 and LL2 maps are comprised of 5 pointings
spaced by 10\farcs6.  We summarize the modules, exposure times, and setups 
used in Table~\ref{tab_IRSsetup}.

All spectra were reduced using the CUBISM software \citep{Smith07} with 
the latest pipeline processing of the data and the most recent calibration 
set (irs\_2009\_05\_20-pb-pfc-trim-omeg-lhllbiasfork.cal).
Each low-resolution module contains two subslits that are exposed at
the same time, e.g., SL1 on-source and SL2 off-source, and vice versa.
We examine the off-source frames, select the ones free of any source 
contamination, trim the maxima and minima, and use the average to produce 
a background frame.  This 2D background frame is then subtracted from all 
on-source frames in the corresponding module to remove astrophysical 
background and to alleviate the bad pixels that contaminate the IRS data.
After the 2D background subtraction, we flag the global and record-level
bad pixels first using the default CUBISM parameters for automatic bad 
pixel detection, and then through manual inspection of each BCD record, 
as well as by backtracking the pixels contributing to a given cube pixel.

The spectra are extracted with aperture sizes large enough to enclose 
the 24 \um\ source. 
See Table 5 for the aperture sizes used for the spectral extractions.
For the LL orders, two local background spectra on either side of the
source are extracted and averaged for a 1D subtraction of the local 
background.  
For the SL orders, a single background spectrum is used for the local 1D 
background subtraction.

Special notes for spectral reduction of individual objects are given below: \\
\underline{WD\,0103+732}.  The frames used for the construction of 
LL1 background contain a very faint point source.  As the WD is bright 
and the point source is practically removed in the min/max trimming and 
averaging process, we choose to ignore this faint source so that we can 
perform the 2D background subtraction to maximize the quality of the final
data cube.
In addition, the frames used for the construction of the LL2 background 
show a small knot of H$_2$ line emission at 17 \um. 
This H$_2$ emission knot is near the center of one slit position, and at the 
edge of the other slit position.  We therefore use only the latter slit 
position for background subtraction.  The pixels of the on-source BCD 
frames affected by over-subtraction of the emission knot are flagged as 
record-level bad pixels, and not used for cube construction. 
Furthermore, this H$_2$ knot is far enough from the WD 
position that both a target spectrum and a local background spectrum can 
be extracted outside the position of the H$_2$ emission knot in the slit. 
The spectrum of WD\,0103+732 is not detected in the SL2 observations.\\
\underline{WD\,0127+581}. This WD is faint, with $F_{24}$ = 0.34 mJy. 
The orientation of the IRS slit, determined by the roll angle of the 
spacecraft, was such that a very bright neighboring star was included 
in the slit.  
Consequently, the weak emission of WD\,0127+581 was overwhelmed by the
elevated background from the neighboring star.
The spectrum of the WD is not unambiguously detected in the 
background-subtracted frame; the signal-to-noise ratio is too low 
for meaningful spectral extraction. \\
\underline{WD\,0439+466}. In its SL2 order, the BCD frames from the three 
final mapping positions show an abrupt jump in brightness across the 
center of each frame.  Since these slit positions do not contain 
the target WD, we do not use them for the SL2 cube construction or 
the SL1 background construction.

\section{Results}  \label{sec:results}

In our MIPS 24 \um\ survey of hot WDs, nine of the 71 objects show 
24 \um\ point sources detected at the given WD positions.
To assess the physical significance of these 24 \um\ detections and 
the 3-$\sigma$ upper limits (see Table 2), we have computed the
expected photospheric emission at 24 \um\ by extrapolating from
the optical and near-IR measurements of each WD.
We first searched the web version of the McCook and Sion 
WD catalog and the SIMBAD database for available 
photometric measurements in the $UBVRIJHK$ bands and the Sloan 
Digital Sky Survey (SDSS) $ugriz$ bands.  For WDs that were 
discovered from the SDSS \citep{Eetal06}, usually only $ugriz$
photometric measurements were available.  We also searched the
2MASS Point Source Catalog \citep{Setal06}
for near-IR counterparts to the WDs.  The effective temperatures of
our program WDs are so high that it is reasonable to approximate their
photospheric emission with a blackbody in the Rayleigh-Jeans limit, i.e.,
the flux density $F_\nu$ $\propto \nu^2$ $\propto \lambda^{-2}$.   
The amounts of extinction to most of these WDs are unknown, but will
have smaller impact at longer wavelengths.
We thus use the photometric measurement with the longest wavelength 
and without extinction correction to extrapolate the expected flux 
density at 23.7 \um, the central wavelength of the MIPS 24 \um\ band.
The fiducial photometric measurement and the expected photospheric 
emission at 24 \um\ are given in columns 6 and 7 of Table 2,
respectively. 

The expected photospheric emission from these hot WDs are all less than
the nominal sensitivity of the 24 \um\ observations, 0.034 mJy.
It is not surprising that all nine detections are at least two orders
of magnitude  higher than the expected photospheric emission, and thus
represent IR excesses.  In the cases of 
non-detections, the 3-$\sigma$ upper limits are several to several tens
times higher than the expected photospheric emission. 
These non-detections do not exclude the existence of low-level 
IR excesses, but no useful deductions can be made from these results; 
thus, the non-detections will not be discussed further in this paper.
We have combined all available photometric measurements of 
the nine WDs and pre-WDs with 24 \um\ excesses to produce the
SEDs plotted in Figure 2.
Below we describe individually the nine hot WDs and pre-WDs with 24 \um\ 
excesses.  Detailed modeling of the SEDs and IRS spectra will be reported
by Bilikova et al.\ (2011b, in preparation).

\subsection{CSPN K\,1-22}

Multi-wavelength optical and IR images of CSPN K\,1-22
are displayed in Figure 3.  The DSS2 Red
image shows a source at 2\arcsec\ north of the CSPN, but no
counterpart of this source is seen in any of the other
optical or IR images.  It is not clear whether this red source 
is spurious or transient.  The {\it HST} images show two sources
separated by 0\farcs35 at the center of K\,1-22; the blue
northeast component is the CSPN and the red southwest 
component is a cool companion \citep{Cetal99}.
The IR images show a source coincident with this close pair 
of stars, but cannot resolve them.

In the SED plot in Figure 2, the flux densities of
CSPN K\,1-22 and its close companion are individually
plotted in the {\it HST} F555W and F814W bands, while
the flux densities in the 2MASS $JHK$ and {\it Spitzer} 
IRAC and MIPS bands are for the two stars combined.
An extinction of $E(B-V) = 0.076$ \citep{Cetal99}
has been corrected from the observed flux densities.
The atmospheric emission of the WD has been modeled as 
a blackbody; for the companion, we adopt the Kurucz 
atmospheric model for a K2\,V star, the spectral type 
implied by the {\it HST} photometry.
These stellar emission models are plotted in thick solid 
curves in Figure 2.
The IRAC and MIPS flux densities are all higher than the 
expected atmospheric emission from these two stars.
The IR excesses in the near-IR and in the mid-IR cannot
be described by a single blackbody; however, two 
blackbody emitters at temperatures of 700 and 150 K 
appear to match the IR excesses in the IRAC and MIPS 24 \um\ 
bands.  The combined SED of the WD, its companion, and 
the two blackbody emitters is plotted as a thin solid line
in Figure 2.

The IRS spectra extracted from CSPN K\,1-22 and adjacent 
background regions all show [\ion{Ne}{3}] 15.55 $\mu$m, 
[\ion{S}{3}] 18.71 \um, [\ion{O}{4}] 25.89 \um, 
and [\ion{S}{3}] 33.48 \um\ line emission (see Figure 4).
The background-subtracted IRS spectrum of CSPN K\,1-22 
shows a weak continuum component and residual line
emission, especially prominent in the [\ion{O}{4}] line.
 
To determine the nature of the residual line emission,
we compare the spatial distributions of the continuum
and line emission.  Figure 5 shows the surface brightness
profiles along the slit for the 24 \um\ continuum, 
[\ion{O}{4}] 25.89 \um\ line, and [\ion{S}{3}] 18.71 \um\ 
line, as extracted from the IRS data cube.   
It is evident that the continuum originates
from a point source, the CSPN, while the line emission from 
the PN is extended.  The [\ion{S}{3}] line has a lower excitation
potential (i.e., ionization potential of S$^+$, 23.3 eV)
and shows an extended, nearly flat surface brightness profile; 
while the [\ion{O}{4}] 25.89 \um\ line has a higher excitation
potential (i.e., ionization potential of O$^{+2}$, 54.9 eV)
and shows a narrower, centrally peaked surface brightness profile.
These different spatial distributions are consistent with the
expectation from the ionization stratification in a PN.

The MIPS 24 \um\ image of K\,1-22 (Figure 1) shows that the 
CSPN is superposed on diffuse emission \citep{Cetal09}.
The IRS spectra show that this diffuse emission is
dominated by the [\ion{O}{4}] 25.89 \um\ line.
The centrally peaked morphology of the diffuse [\ion{O}{4}]
emission makes it difficult to subtract the background accurately.
As the background spectrum is the average of those extracted 
from the two regions at $\pm$20$''$ from the central source, 
it can be seen from the [\ion{O}{4}] surface brightness profile 
in Figure 5 that the excess emission at the center is $\sim$35\% 
of the background emission.  The residual [\ion{O}{4}] emission 
in the background-subtracted spectrum in Figure 4 is indeed 
about 1/3 as strong as the emission from the background regions.
Therefore, the apparent [\ion{O}{4}] emission in the 
background-subtracted spectrum of CSPN K\,1-22 is most likely a 
residual from imperfect background subtraction.

Figure 4 shows that the MIPS 24 \um\ band photometric 
measurement ($\sim$ 1 mJy)  of the CSPN, plotted as an open diamond, appears 
higher than the continuum flux density of the 
background-subtracted spectrum ($\sim$0.75 mJy).  
This discrepancy is likely caused by the contamination
of [\ion{O}{4}] line in the MIPS 24 \um\ band.
The photometric measurements in the IRAC 5.8 
and 8.0 \um\ bands, plotted as open diamonds in Figure 4, 
are also higher than the continuum flux densities; however,
these discrepancies are less than 25\% and are reasonable for the
S/N of the spectrum at these wavelengths.

The background-subtracted IRS spectrum of CSPN K\,1-22 shows
continuum emission well above the expected photospheric emission
of the CSPN and its red companion ($\sim$0.011 mJy), and thus 
represents an IR excess due to dust continuum.  
Nevertheless, it is not known whether the excess IR emission 
is associated with the CSPN or its red companion.
High-resolution images in the mid-IR wavelengths are needed to 
resolve CSPN K\,1-22 and its companion.

\subsection{CSPN NGC\,2438}

CSPN NGC\,2438 shows bright 24 \um\ emission more than four 
orders of magnitude higher than the expected photospheric emission.
The MIPS observation of NGC\,2438 was made after the {\it Spitzer} 
Cycle 5 proposal deadline; thus NGC\,2438 was not an IRS target in
our Program 50629.  However, \citet{Betal09} had found IR excesses 
of CSPN NGC\,2438 through the analysis of archival IRAC observations
(see IRAC images in Figure 6), and included NGC\,2438 in another 
{\it Spitzer} program to study IR excesses of CSPNs (Program 50793;
PI: Bilikova).  The IRS observations used only the short wavelength
modules and therefore only extend to $\sim$15 \um.  
The background-subtracted IRS spectrum of the CSPN NGC\,2438 exhibits 
continuum emission (Bilikova et al.\ 2011b, in preparation).  

The SED of CSPN NGC\,2438 (Figure 2) shows that the 24 \um\ excess 
is much greater than the IR excesses in the IRAC bands. 
The IR excesses in the IRAC bands and the MIPS 24 \um\ band cannot
originate from a single-temperature emitter; instead, they can be
roughly described by two emitters at 1200 K and 150 K, respectively.
Unfortunately, 2MASS did not detect the CSPN NGC\,2438 in the $H$ 
and $K$ bands to allow a more precise modeling of the SED to 
determine the temperature of the hotter emitter and whether it 
originates from a low-mass stellar companion or from a dust disk.

\subsection{WD\,0103+732 (CSPN EGB\,1)}

WD\,0103+732 is the central star of the PN EGB\,1.  {\it HST} 
images do not show any companion stars \citep{Cetal99}.  
See Figure 7 for optical, 2MASS $J$, IRAC, and MIPS 24 $\mu$m 
images of WD\,0103+732.  The SED of WD\,0103+732 (Figure 2)
shows optical and near-IR flux densities following a blackbody 
curve closely, although the WD is not detected in the 2MASS $H$ 
and $K$ bands, where 2-$\sigma$ upper limits are plotted.  The 
flux density in the IRAC 8.0 \um\ band starts to rise above the
photospheric emission level, and in the MIPS 24 \um\ band the
flux density is more than three orders of magnitude higher than
expected from photospheric emission.

The MIPS 24 \um\ image (Figure 1) shows that the point source of 
WD\,0103+732 is superposed on diffuse emission \citep{Cetal09}.
The IRS spectrum of the background  (see Figure 8) shows that the 
diffuse emission is dominated by line emission with a weak but 
appreciable continuum at wavelengths $\gtrsim$ 20 \um, and that 
the major contributor to the MIPS 24 \um\ emission is the 
[\ion{O}{4}] 25.89 \um\ line.
The background-subtracted spectrum of WD\,0103+732 (Figure 8) is dominated 
by continuum emission.  The apparent deficit at the [\ion{Ne}{3}] 
and [\ion{S}{3}] lines and excess at the [\ion{O}{4}] line are
caused by imperfect background subtraction. 
Figure 9 shows the surface brightness profiles of the 24 $\mu$m
continuum, [\ion{O}{4}] and [\ion{S}{3}] lines extracted from the
IRS data cube.  The nebular emission near WD\,0103+732
also shows ionization/excitation stratification, as the 
[\ion{O}{4}] emission profile is narrower and more strongly peaked 
toward the center than [\ion{S}{3}], resulting in an under-subtraction
of background in the [\ion{O}{4}] line and over-subtractions in
the [\ion{S}{3}] and [\ion{Ne}{3}] lines.
Interestingly, the background shows an emission feature at 17 \um,
which could be associated with transitions of H$_2$ or polycyclic 
aromatic hydrocarbons (PAH).  As the emission feature is narrow
and the PAH 11.2 \um\ feature is not present, we identify this 
17 \um\ feature as the H$_2$ (0, 0) S(1) line emission.
This H$_2$ emission is associated with a much more extended background
and does not vary significantly over the regions of spectral extractions,
no residual H$_2$ 17 \um\ line emission is present in the 
background-subtracted spectrum of WD\,0103+732.

Despite the morphological differences between the PNe EGB1 and the Helix
Nebula, the SED and IRS spectrum of WD\,0103+732 are very similar to
those of WD\,2226$-$210 \citep{Suetal07}.  Both show IR excess starting 
in the IRAC 8 $\mu$m band and peaking near the MIPS 70 $\mu$m band.
Because of the bright nebulosity and lower angular resolution of the
MIPS 70 $\mu$m camera, we consider the 70 \um\ flux density of 
WD\,0103+732, 55 mJy, an upper limit. 
As the effective temperature of WD\,0103+732, $\sim$150,000 K, is higher
than that of WD\,2226$-$210, $\sim$110,000 K, the physical properties
of their dust disks are different (Bilikova et al.\ 2011b, in preparation).
Note that the light curve of WD\,0103+732 shows sinusoidal variations in the 
$BVR$ bands, but the nature of these variations is uncertain (Hillwig et 
al.\ 2011, in preparation).

\subsection{WD\,0109+111}

This WD is not surrounded by any known PN.  It is not in a crowded
region.  The optical and 2MASS sources are coincident within 1\arcsec;
the 24 \um\ source is faint, but also coincident with the optical
source within 1-2\arcsec\ (see Figure 10).  MIPS 24 \um\ images are usually 
infested with faint background galaxies.  We identified $\sim$20 sources 
within a 3$'\times3'$ area centered on WD\,0109+111.  If these sources are
randomly distributed within this area, the probability to have one
source landing within 2\arcsec\ from WD\,0109+111 is $4\times10^{-4}$.
Thus, it is unlikely that the 24 \um\ source is a chance superposition
of a background source.  

The SED of WD\,0109+111 (see Figure 2) shows that the optical and
near-IR flux densities fall nicely along a blackbody curve, but
the 24 \um\ emission is more than 100 times higher than the expected
photospheric emission.  WD\,0109+111 is the second faintest 24 \um\ source
among the nine detections.  No IRS spectra are available for WD\,0109+111.

\subsection{WD\,0127+581 (CSPN Sh\,2-188)}

WD\,0127+581 is surrounded by the PN Sh\,2-188.  This WD is not
detected in the 2MASS $JHK$ bands.  We have thus used the Kitt Peak 
National Observatory 2.1 m telescope with the FLAMINGOS detector and
measured $J$ = 17.03$\pm$0.13 and $K_s$ = 16.18$\pm$0.13.
We have also obtained IRAC observations of this object.  See
Figure 11 for these images of WD\,0127+581.
As WD\,0127+581 is a faint source superposed on a bright background,
the errors in photometric measurements are large. 
The WD is easily visible in the 3.6 and 4.5 \um\ bands, but not 
convincingly detected in the 5.8 and 8.0 \um\ bands. 
As explained in Section 2.3, the weak emission from WD\,0127+581
is superposed on an elevated background from a bright neighboring
star in the IRS observations.  The high noise level prohibits the
detection of the spectra of WD\,0127+581, and hence no spectra are 
extracted.

The SED of WD\,0127+581 (Figure 2) is complex.  The near-IR excess
in the $J$ and $K$ bands is indicative of a late-type companion, while
the IR excesses in the IRAC bands and MIPS 24 \um\ band require two 
emitters at different temperatures.  Two blackbody emitters at
temperatures of 900 K and 150 K are plotted in Figure 2 to illustrate
one possibility.

\subsection{WD\,0439+466 (CSPN Sh\,2-216)}

WD\,0439+466 is the central star of the PN Sh\,2-216 at a distance 
of 129$\pm$6 pc \citep{Hetal07}.
The WD is coincident with a point source surrounded by diffuse emission
in the MIPS 24 \um\ image \citep{Cetal09}.  The SED of WD\,0439+466
follows the photospheric blackbody curve throughout the optical, 2MASS, 
and IRAC bands, but shows a large deviation in the MIPS 24 \um\ band.  
See Figure 12 for optical and IR images of WD\,0439+466.

The IRS spectra of WD\,0439+466 and its background are shown in Figure 13.
The spectrum of the diffuse emission adjacent to WD\,0439+466 shows the
[\ion{O}{4}] 25.89 \um\ and the H$_2$ (0, 0) S(1) 17 \um\ lines, in 
addition to a weak continuum that is appreciable at wavelengths greater 
than $\sim$25 \um.
The spectrum extracted at the position of WD\,0439+466 also shows the
[\ion{O}{4}] and H$_2$ lines, but these emission features 
are effectively removed by the subtraction of the local nebular background.
The background-subtracted spectrum of WD\,0439+466 is totally dominated by 
continuum emission in the 15--35 \um\ range; furthermore, its flux 
densities agree well with the photometric measurements in the IRAC 5.8 and 
8.0 \um\ and MIPS 24 \um\ bands. The continuum-dominated nature of
this spectrum is similar to that of the Helix central star.  The detailed
modeling of WD\,0439+466's SED and spectrum will be presented by
Bilikova et al.\ (2011b, in preparation).

\subsection{WD\,0726+133 (CSPN Abell 21)}

WD\,0726+133 is the central star of the PN Abell 21, also known as YM\,29.
The WD appears as a point source superposed on diffuse emission in the MIPS
24 \um\ image \citep{Cetal09}.  See Figure 14 for optical and IR images
of WD\,0726+133.  The flux densities of WD\,0726+133 in
the optical, 2MASS $JHK$, and IRAC bands are all consistent with a 
blackbody approximation of the stellar photospheric emission.  
The flux density in the MIPS 24 \um\ band is almost three orders of 
magnitude higher than the expected photospheric emission.  
Several 24 \um\ sources are detected near WD\,0726+133, and their optical
counterparts are resolved into spiral galaxies in archival {\it HST} 
F555W and F814W images, but WD\,0726+133 remains unresolved and shows no
companions \citep{Cetal99}.  No IRS spectra are available for this WD.

\subsection{WD\,0950+139 (CSPN EGB\,6)}

WD\,0950+139 is surrounded by the PN EGB\,6.  This WD previously gained 
attention because of its strong [\ion{O}{3}] and [\ion{Ne}{3}] emission 
lines \citep{Letal89} and near-IR excesses \citep{FL93}.
This puzzle was partially solved by an [\ion{O}{3}] image taken with
the {\it HST} Faint Object Camera, which revealed an unresolved source
0\farcs18 from the WD \citep{Bond94,Bond09}.  
Apparently, WD\,0950+139 has a late-type companion and the line 
emission is associated with the companion.
See Figure 15 for optical and IR images of WD\,0950+139.
Our MIPS 24 \um\ observation of WD\,0950+139 shows a bright unresolved
source, and the 24 \um\ flux density is four orders of magnitude higher
than the photospheric emission expected from WD\,0950+139 (Figure 2).
The SED in the IRAC and MIPS 24 \um\ bands cannot be accounted for by a
single-temperature emitter; instead, it may be described by two emitters 
at temperatures of 500 and 150 K.
A {\it Spitzer} IRS spectrum of WD\,0950+139 was obtained
through Guaranteed Time Observations, and the spectrum confirmed that
the emission in the MIPS 24 \um\ band is dominated by continuum
(Su et al.\ 2011, in preparation).

\subsection{WD\,1342+443}

This WD was discovered in the SDSS.  It is below the detection
limit of 2MASS.  Figure 16 shows optical and IR images of WD\,1342+443.
The SED of WD\,1342+443 in Figure 2 shows the SDSS photometric
measurements falling along the blackbody model curve with the
24 \um\ flux density more than 400 times higher than the expected
photospheric emission.  This WD has the faintest 24 \um\ flux
density among the nine hot WDs detected.  No IRS spectra are 
available for WD\,1342+443.

\section{Discussion}  \label{sec:discussion}

\subsection{Statistical Properties}

Among our sample of 71 hot WDs, nine show 24 \um\ excesses,
corresponding to almost 13\%.  Figure 17 shows the distribution
of the sample in $V$ and $J$.  The number of detections in each
magnitude bin is too small to provide meaningful statistics.  
If the sample is divided into a brighter group that has 
photometric measurements and a fainter group that has no 
photometric measurements, it can be seen that 24 \um\ excesses
are detected in 15-16\% of the brighter hot WDs, and only
$\sim$8\% among the fainter hot WDs.  As all of the WDs in our 
sample have high temperatures and a small range of radii, their 
brightnesses are indicative of distances, with the fainter ones 
being at larger distances.  The different percentages of brighter 
and fainter WDs exhibiting 24 \um\ excesses most likely reflect 
the fact that the limiting sensitivity of our MIPS 24 \um\ survey 
precludes the detection of distant objects.  The true percentage of hot 
WDs exhibiting 24 \um\ excesses is likely greater than 15-16\%.

For the nine hot WDs with 24 \um\ excesses, their 24 \um\ flux
densities are plotted against their $J$-band flux densities in
Figure 18.  No correlations are seen in this plot.  This is
expected, as the $J$-band flux density is a rough indicator
of distance and the 24 \um\ excess should not be dependent on
distance.  Another reason for the lack of correlation is the
diverse physical conditions of the 24 \um\ emitters, as discussed
later in Section 4.2.

Seven of the nine hot WDs with 24 \um\ excesses are still surrounded
by PNe: CSPN K\,1-22, CSPN NGC\,2438, WD\,0103+732 in EGB\,1, 
WD\,0127+581 in Sh\,2-188, WD\,0439+466 in Sh\,2-216, 
WD\,0726+133 in Abell\,21 (YM\,29),  and WD\,0950+139 in EGB\,6;
while two are not in PNe: WD\,0109+111 and WD\,1342+443.
There is a striking difference in the frequency of occurrence of
24 \um\ excesses between the WDs and pre-WDs in PNe and those
without PNe: 20\% and 5--6\%, respectively.
The two WDs not in known PNe are also the faintest in 24 \um\ among the
nine.   WDs without PNe are more evolved than those that are still 
surrounded by PNe.  The smaller 24 \um\ excesses of the WDs without
PNe appear to indicate an evolutionary trend of diminishing excess;
however, this trend is not obvious among the 24 \um\ excesses of the
WDs with PNe, if the nebular sizes \citep{Cetal09} are indicative of 
their evolutionary status.  Considering the diversity in the progenitors' 
masses and evolutionary paths of these WDs, the current sample of hot WDs
with 24 \um\ excesses is too small to allow us to distinguish between
an evolutionary effect and other effects.

\subsection{Nature of the 24 \um\ Excesses}

The IR excesses of the CSPN Helix are detected in the 8, 24, and 70 \um\ bands,
but not at shorter wavelengths \citep{Suetal07}. High spatial
resolution {\it HST} observation has ruled out any resolved companion earlier
than M5 \citep{Cetal99}. Furthermore, the photometric accuracies in the
IRAC bands (1-$\sigma$ of 20, 24, 26 and 17 $\mu$Jy at the 3.6, 4.5, 5.4,
8.0 $\mu$m bands, respectively) can further constrain the mass of a
possible companion. At an age of 1 Gyr, the 2MASS and IRAC photometry
can also rule out, at the 3-$\sigma$ level, any companion with mass greater
than 20 Jupiter masses, i.e., early T dwarfs with $T_{\rm eff} 
\lesssim$650 K.
Three of the nine new detections of mid-IR excesses of hot WDs show 
similar SEDs: WD\,0103+732 (CSPN EGB\,1) shows excess emission at 
8 and 24 \um, while WD\,0439+466 (CSPN Sh\,2-216) and WD\,0726+133
(CSPN Abell 21) show excess emission at only 24 \um.
Their lack of near-IR excesses does not support the presence of late-type
companions.

Four of the hot WDs with 24 \um\ excesses also exhibit excess emission
in the IRAC bands: CSPN K\,1-22, CSPN NGC\,2438, WD\,0127+581 
(CSPN Sh\,2-188), and WD\,0950+139 (CSPN EGB 6).  IRS observations
indicate that this IR excess is continuum in nature, and crude fits
to the SEDs suggest blackbody emitter temperatures of 500--1200 K.
These temperatures are in the ranges for brown dwarfs, but the
emitting areas, $\sim 1\times10^{23}$ to $3\times10^{25}$ cm$^2$,
are too large for brown dwarfs.  
It is most likely that these IR excesses in the IRAC bands originate
from hot dust emission.
The two cases with cooler emitter temperatures (500-700 K), CSPN K\,1-22 
and WD\,0950+139, are known to have late-type companions 
\citep{Cetal99,Bond09}.
The relationship between these companions and the hot dust components
is uncertain.
Future searches for companions of CSPN NGC\,2438 and WD\,0127+581
may help us understand the roles played by stellar companions in
producing the dust component responsible for the IRAC excesses.
Finally, two hot WDs with 24 \um\ excesses, WD\,0109+111 and 
WD\,1342+443, have no IRAC observations to determine whether excess 
emission is present in the IRAC bands; their 2MASS measurements do 
show any excess emission.
In summary, among the ten hot WDs and pre-WDs that show excess 
emission at 24 \um, 40\% also show near-IR excesses associated with 
an additional warmer emission component that might be related to the
presence of a companion.

Whether a 24 \um\ excess is accompanied by near-IR excess
or not, the shape of the SED suggests that the 24 \um\
emission originates from a component distinct from the 
WD's photospheric emission or another near-IR emitter.
The 24 \um\ emission must originate from a source cooler
than 300 K.  Assuming that this source is heated solely
by stellar radiation, we can determine the covering factor
of the emitter from the luminosity ratio $L_{\rm IR}/L_*$, where
$L_{\rm IR}$ is the luminosity of the excess emitter and 
$L_*$ is the luminosity of the WD.  We adopt the stellar 
effective temperature, assume a blackbody model, and use the
distance and extinction corrected photometry to determine
the stellar luminosity $L_*$. (Useful physical parameters 
are listed in Table 6.)
The luminosity of the excess emitter is calculated by
assuming a 150 K blackbody model normalized at the 
observed 24 \um\ flux density. In the case of WD\,0103+732 
and WD\,2226$-$210, blackbody temperatures are determined
from model fits to the 8, 24, and 70 \um\ flux densities.  
For the four WDs exhibiting excess emission in the IRAC bands,
we add another blackbody component to fit these measurements. 
These IR emitter models, illustrated in Figure 2, 
are used to calculate their approximate luminosities.
The $L_{\rm IR}/L_*$ ratios of the nine new cases and the 
Helix CSPN are listed in Table 6.
The covering factors range from $4.9\times10^{-6}$ to 
$4.7\times10^{-4}$. 

A perfect absorbing body heated by a 100,000 K WD to 150 K 
would be at a distance of $\sim$10 AU, and 100 K at 20 AU.
Even at a distance of 10 AU, a covering factor of 
$4\times10^{-6}$ corresponds to $5\times10^{-3}$ AU$^2$,
or $3\times10^6$ $R_\odot^2$, too large to be a star or
a planet.  The most likely origin of this mid-IR emitter is
a dust disk, as proposed for the CSPN Helix Nebula 
\citep{Suetal07}.
The presence of near-IR excesses indicates the existence
of a binary companion or a dust disk at higher temperatures, or
both.  Detailed modeling and high-resolution images are needed
to decipher the true nature of the IR excesses.  We defer the
modeling of the SEDs and IRS spectra of hot WDs with 24 \um\
excesses to another paper (Bilikova et al.\ 2011b, in preparation).

The connection among CSPNs, circumstellar dust disks, and binarity
has been alluded to by various observations and theoretical models.
For example, Keplerian circumstellar dust disks have been observed
to be associated with binary post-AGB stars \citep{DRetal06}.
Furthermore, the close binary stellar evolution of CSPNs has been
suggested to play a very important role in the formation and shaping 
of PNe \citep{Soker98,NB06,dM09}, although such binary companions are 
difficult to detect.
Our detection of a warmer emitter (500--1200 K) in addition to 
the colder dust component (at 100--200 K) in four CSPNs, two of which 
have known companions, appear to further the connection.  However, 
the dust disks responsible for 24 \um\ excesses are much larger than
the circumstellar disks ejected from common-envelope binaries
\citep[a few AU at most;][]{TR10}, and have very different geometry from the 
Keplerian circumstellar dust disks around binary post-AGB stars, 
especially in the covering factor ($L_{\rm IR}/L_*$).
The Keplerian circumstellar dust disks around binary post-AGB stars
typically have $L_{\rm IR}/L_*$ $\sim$ 0.2--0.5 \citep{DRetal06}, several
orders of magnitude higher than those responsible for 24 \um\ excesses 
of hot WDs and pre-WDs.
This large discrepancy in covering factors suggests that these
dust disks have different origins, and the small covering factors
are more consistent with debris disks observed in main sequence
stars \citep{Suetal06,Tetal08,wyatt08}.
It is thus likely that the origin of the 24 \um\ excesses of hot WDs
and pre-WDs is dynamically rejuvenated debris disks as suggested by
\citet{Suetal07}.

\section{Summary}

The central WD of the Helix Nebula has been shown to exhibit
excess emission in the {\it Spitzer} MIPS 24 and 70 \um\ bands
and it is suggested that this IR excess originates from a dust disk 
produced by collisions among KBOs \citep{Suetal07}. 
Inspired by the WD in the Helix, we have conducted a MIPS 24 \um\ 
survey of 71 hot WDs and pre-WDs and found excess 24 \um\ emission 
for nine of them.  We have further obtained {\it Spitzer} IRAC and IRS
follow-up observations for a subset of these WDs with 24 \um\ excesses.

The detection of 24 \um\ excesses is limited by the sensitivity
of the observations.  Among the hot WDs with optical or near-IR
photometric measurements, 15-16\% are detected at 24 \um\ with
excesses.  The true occurrence rate of 24 \um\ excesses must be
still higher.  Thirty-five of the 71 WDs and pre-WDs in our survey
sample are in known PNe; among these 20\% show 24 \um\ excesses,
while among the 36 WDs without known PNe only 5--6\% show 24 \um\
excesses.  

The 24 \um\ excesses are accompanied by different levels of
excesses in the $JHK$ and IRAC bands.
The excess emission in the $JHK$ bands for CSPN K\,1-22 and 
WD\,0950+139 (CSPN EGB\,6) originates from their known low-mass 
companions.  Excess emission in the IRAC bands are present in 
CSPN K\,1-22, CSPN NGC\,2438, WD\,0950+139, and WD\,0127+581 
(CSPN Sh\,2-188); IRS spectra show continuum emission clearly
in the three former objects and faintly in WD\,0127+581,
indicating the presence of a dust component at temperatures
of $\sim$1000 K.
Only WD\,0103+732 shows 8, 24, and 70 \um\ excesses similar 
to those in the SED of the Helix central WD.
In two cases, WD\,0439+466 and WD\,0726+133, no IR excess is
present in the IRAC bands; the IRS spectrum of WD\,0439+466 shows 
that its 24 \um\ band emission is dominated by dust continuum.

The emitters responsible for the 24 \um\ excesses have large 
emitting surface areas that can be provided only by dust disks.
Furthermore, the excess emission in these long wavelength bands
indicates dust temperatures $<$300 K.
The $L_{\rm IR}/L_*$ ratios of dust disks responsible for the
24 \um\ excesses of hot WDs and pre-WDs are in the range of 
$4.9\times10^{-6}$ to $4.7\times10^{-4}$, similar to those 
observed in debris disks around main sequence stars, but 
several orders of magnitude lower that those of Keplerian 
circumstellar dust disks around binary post-AGB stars.
It is likely that these dust disks around hot WDs and pre-WDs
are indeed rejuvenated debris disks as suggested by \citet{Suetal07}
for the Helix central star.

In two cases, CSPN K\,1-22 and WD\,0950+139 in EGB\,6,
late-type companions have been resolved by {\it HST} images,
but {\it Spitzer} images could not resolve these CSPNs and their
companions.  It is thus uncertain whether the dust disks are
associated with the CSPNs or their companion.  High-resolution
mid-IR images are needed to establish the associations unambiguously.

\acknowledgments
This research was supported by the NASA grants JPL 1319342
and 1343946 and SAO GO8-9026.  
M.A.G.\ acknowledges support from the Spanish Ministerio de 
Ciencia e Innovaci\'on (MICINN) through grant AYA2008-01934.
We thank George Rieke for useful discussion, Adam Myers for 
advice on the SDSS photometry, Adeline Caulet and Ian McNabb
for assisting in the preliminary IRS data reduction, and the
anonymous referee and Jay Farihi for suggestions to improve the paper.
The Digitized Sky Survey images used were produced at the Space Telescope 
Science Institute under U.S. Government grant NAG W-2166. The images of 
these surveys are based on photographic data obtained using the UK Schmidt 
Telescope. The plates were processed into the present compressed digital 
form with the permission of these institutions.
Images from the NASA/ESA Hubble Space Telescope were obtained from the 
data archive at the Space Telescope Science Institute.  STScI is operated 
by the Association of Universities for Research in Astronomy, Inc. under 
NASA contract NAS 5-26555.
This research has made use of 
the SIMBAD database, operated at CDS, Strasbourg, France.
This research has also made use of the NASA/IPAC Infrared 
Science Archive, which is operated by the Jet Propulsion 
Laboratory, California Institute of Technology, under contract 
with the National Aeronautics and Space Administration.


\begin{figure}  
\epsscale{0.7}
\plotone{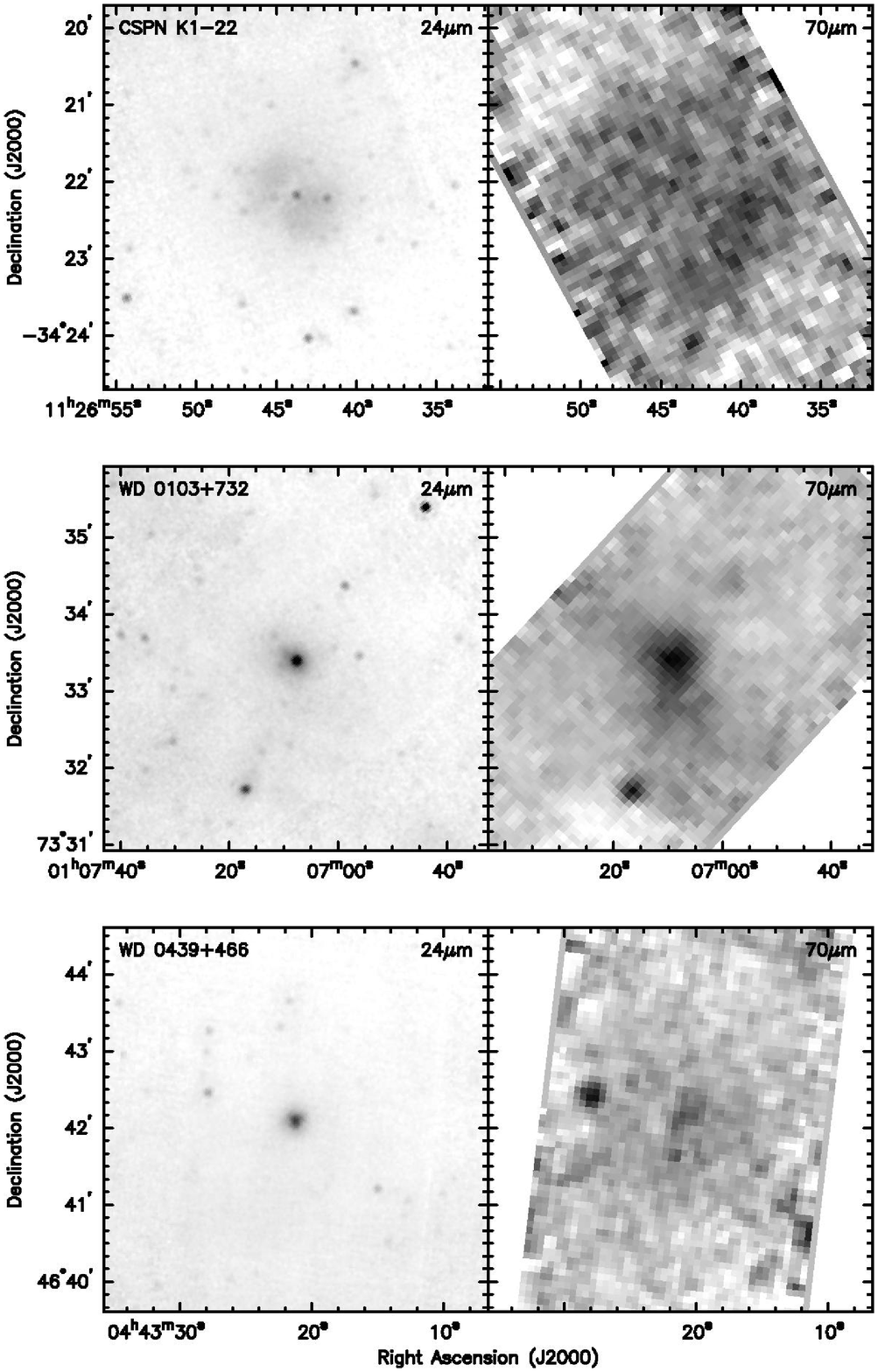}
\caption{{\it Spitzer} MIPS 24 and 70 \um\ images of CSPN K\,1-22, 
WD\,0103+732 (CSPN EGB\,1), and WD\,0439+466 (CSPN Sh\,2-216).}
\end{figure}

\begin{figure}  
\epsscale{0.85}
\plottwo{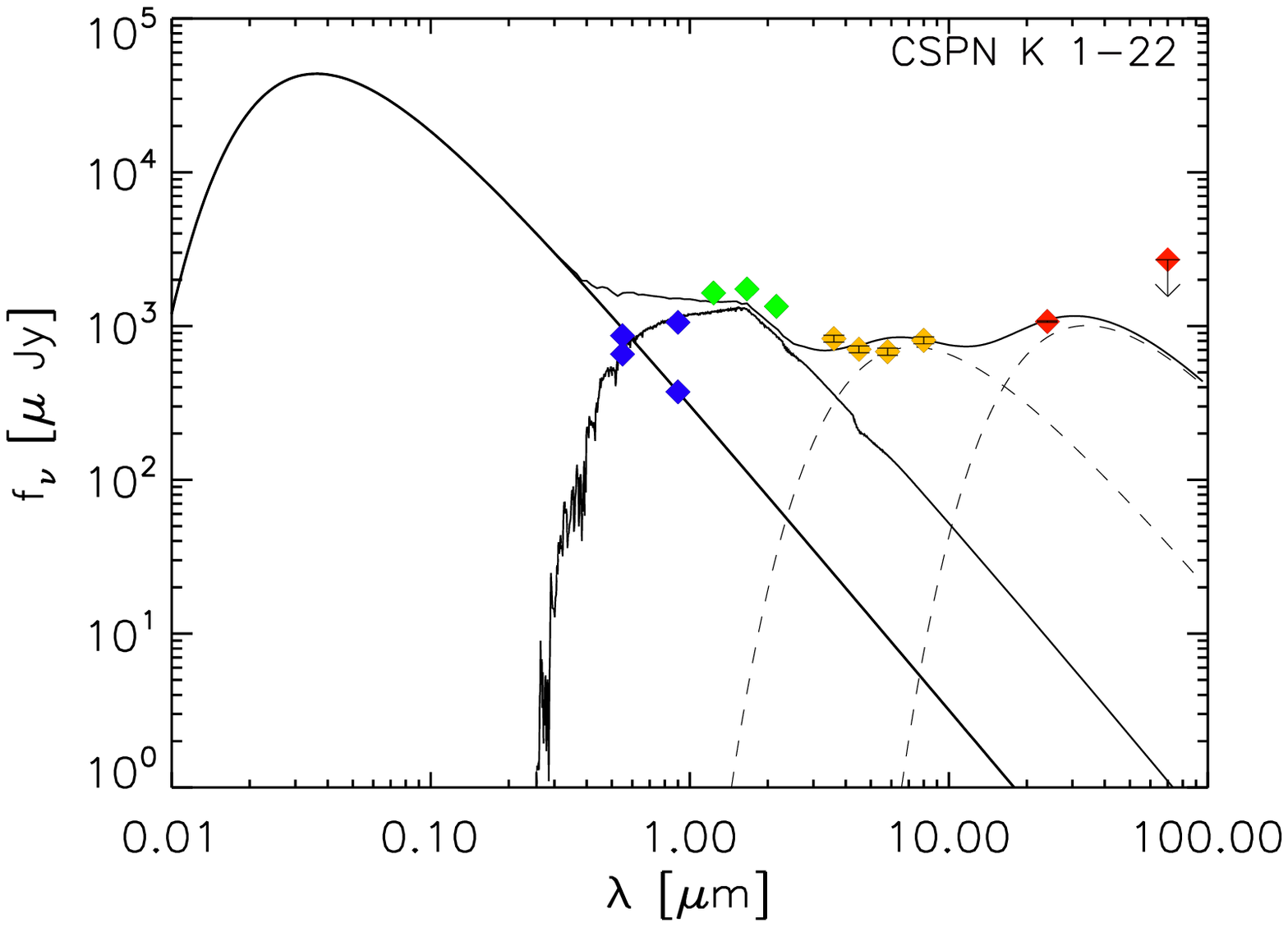}{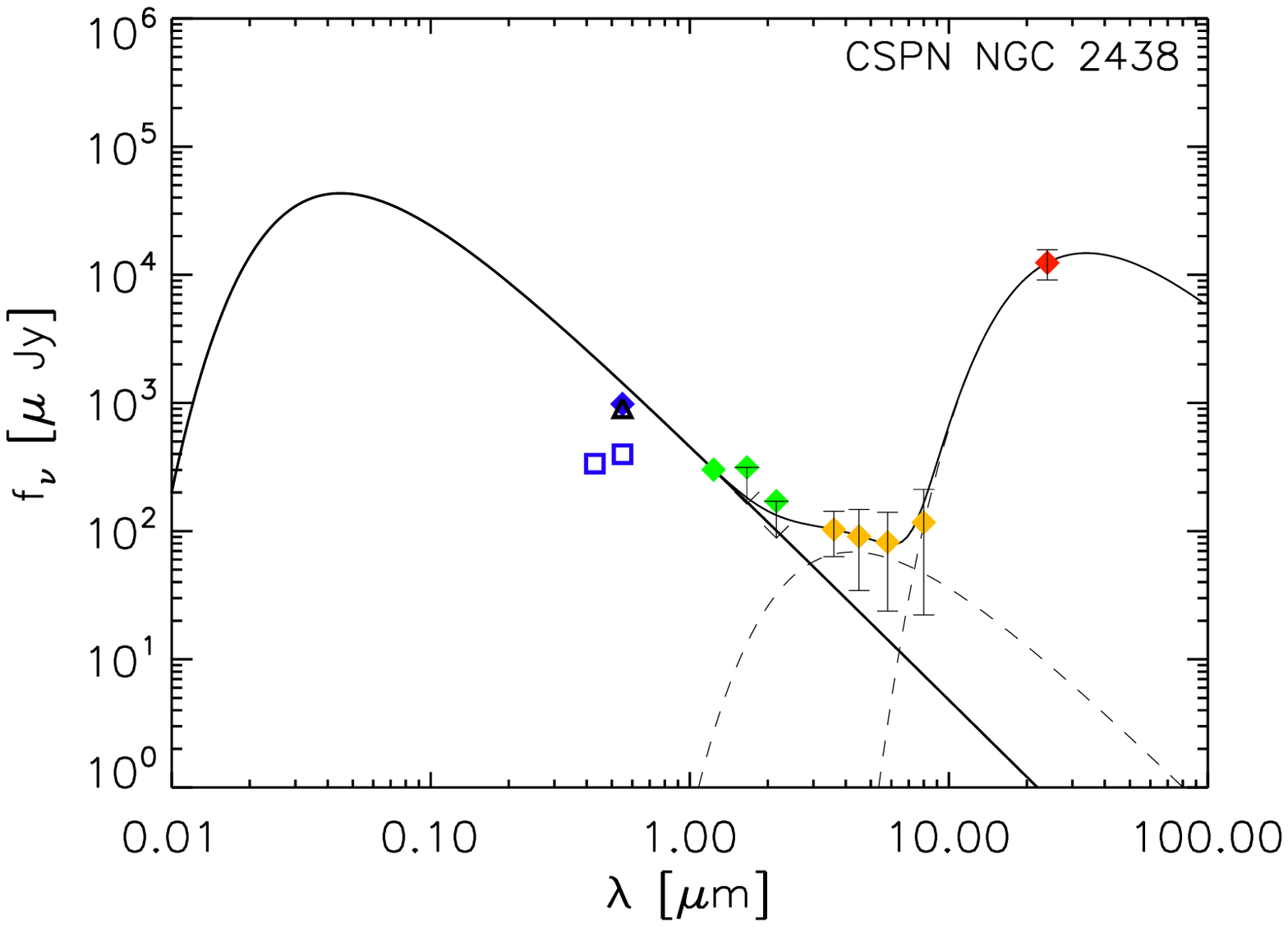} \\
\plottwo{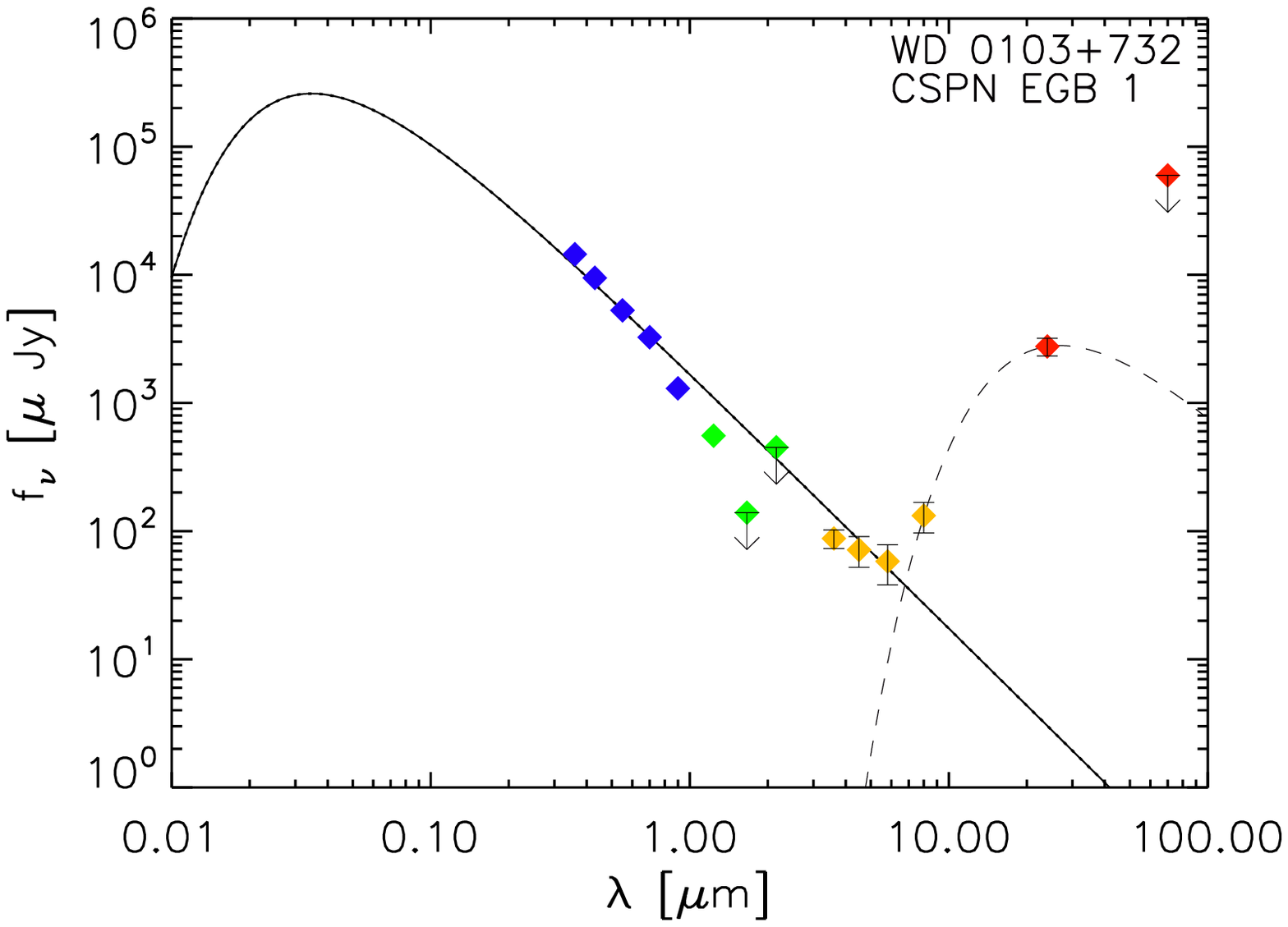}{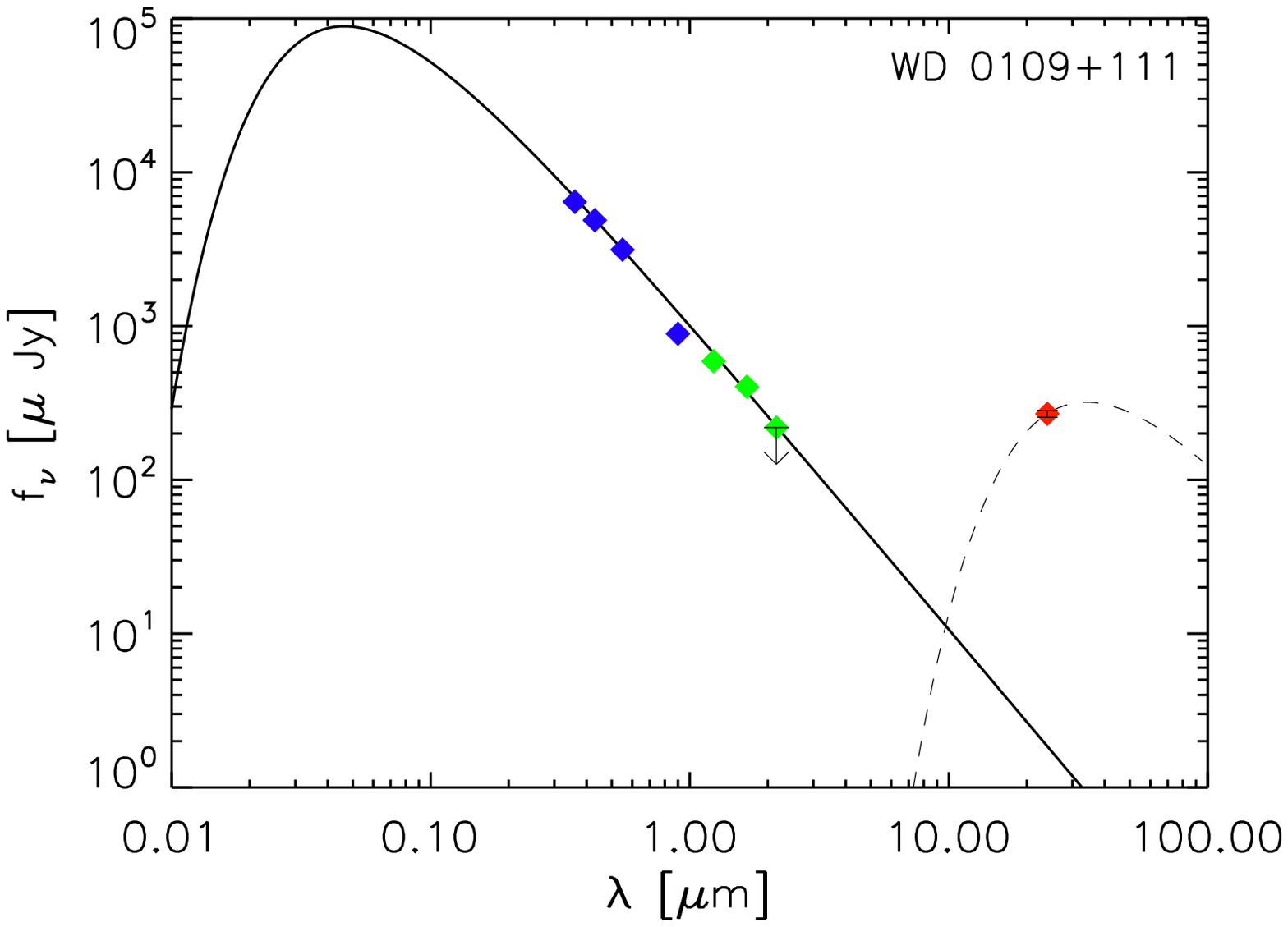}\\
\plottwo{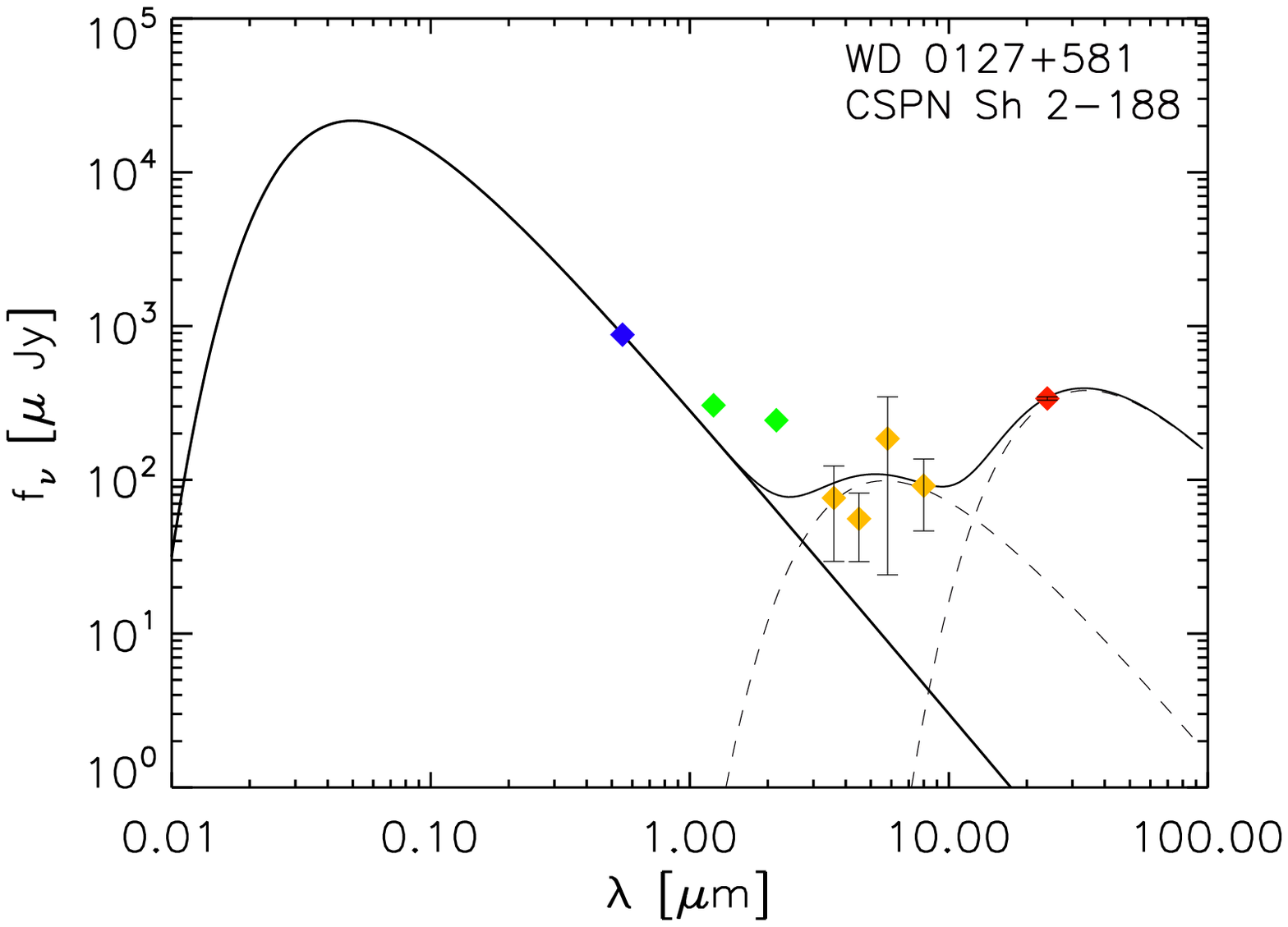}{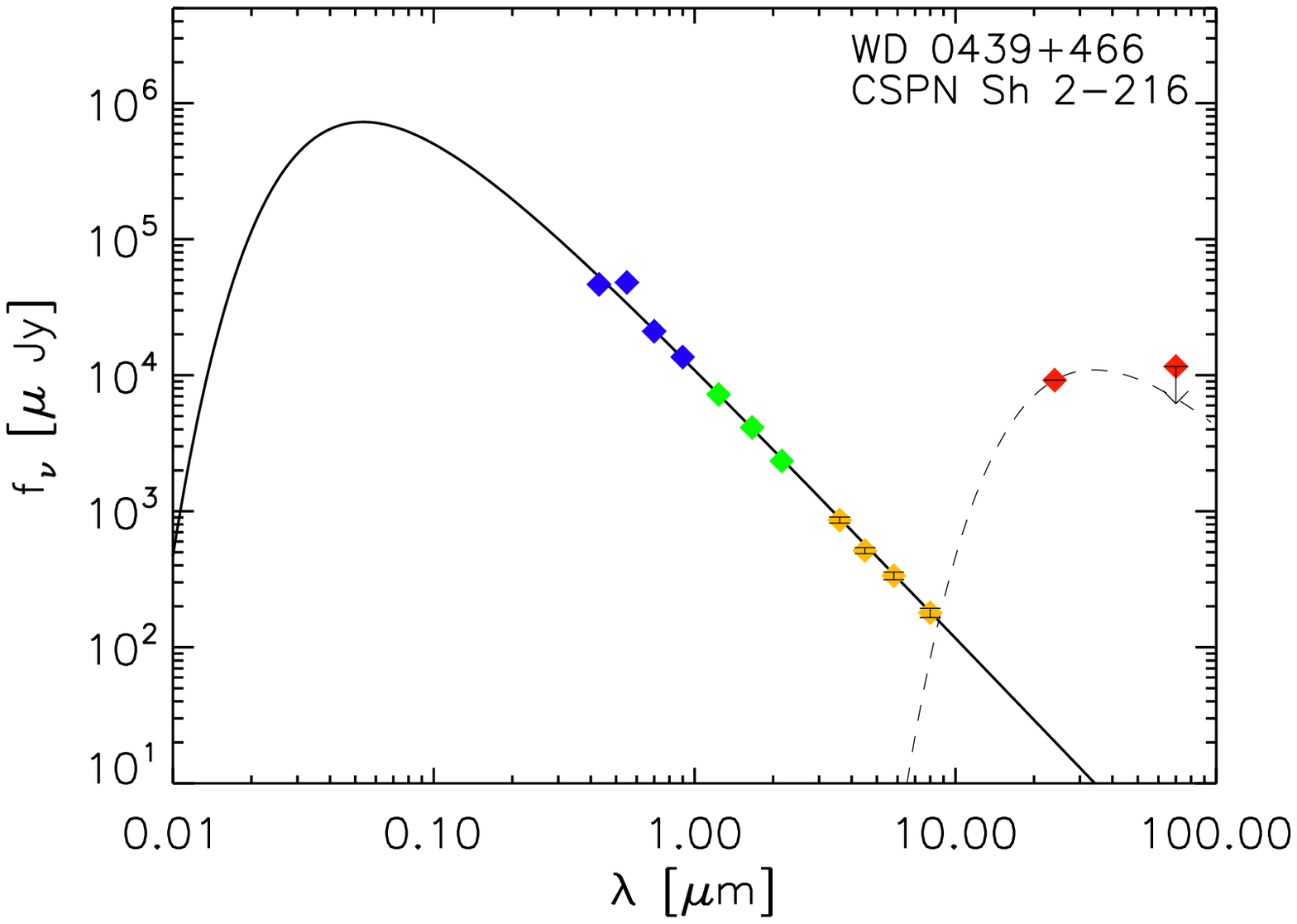}
\end{figure}
\begin{figure}
\epsscale{0.85}
\plottwo{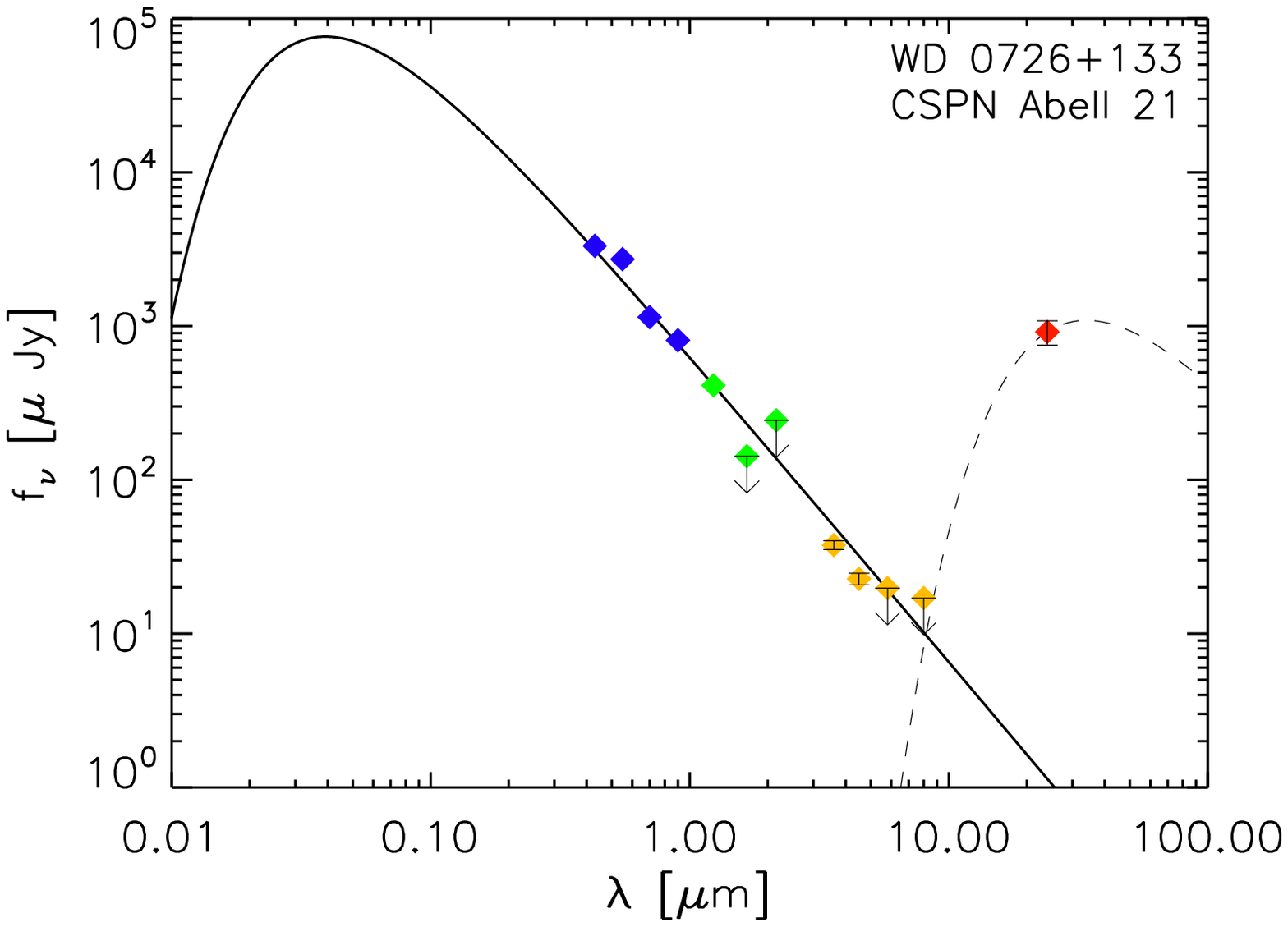}{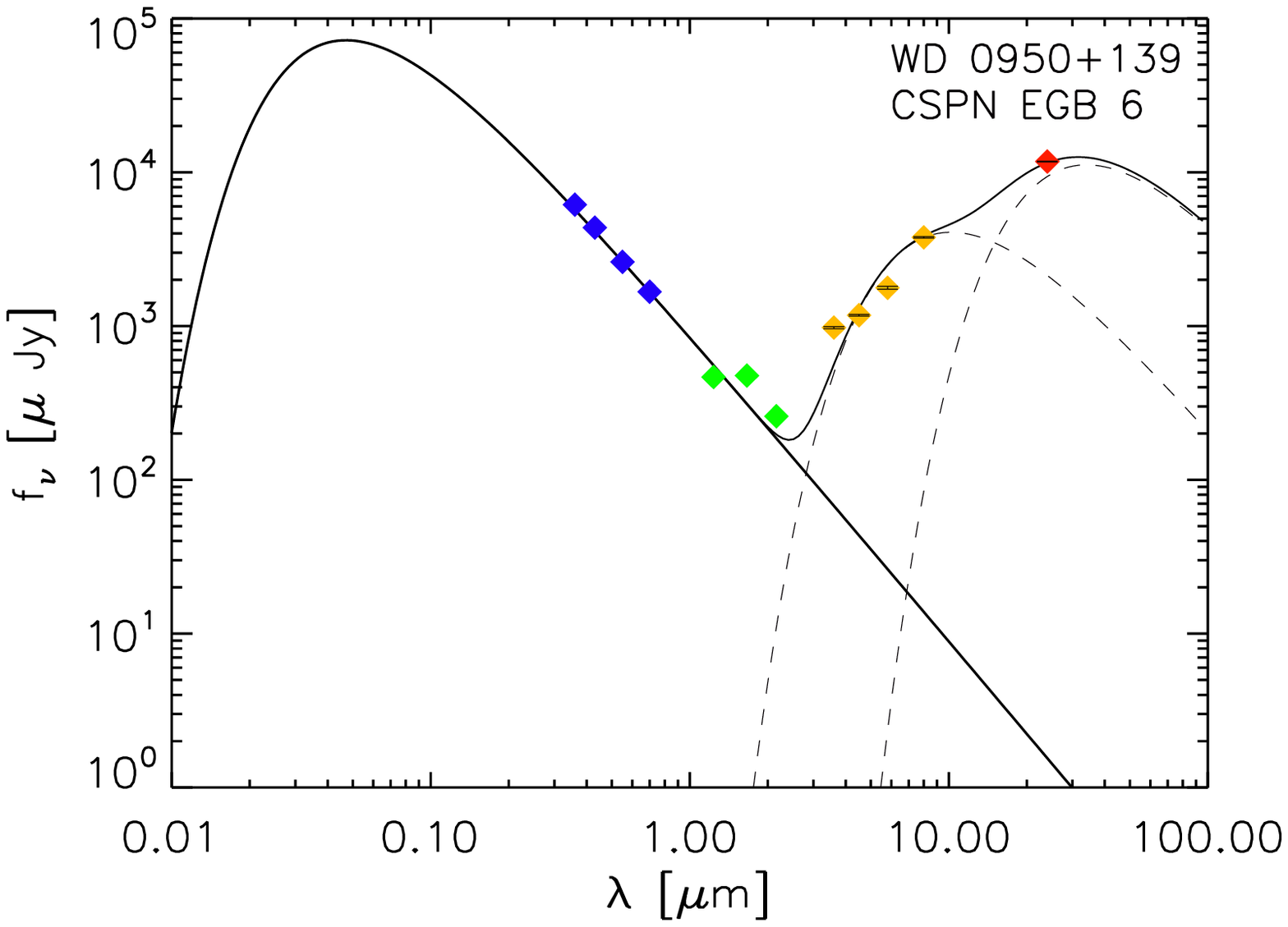} \\
\epsscale{0.4}
\plotone{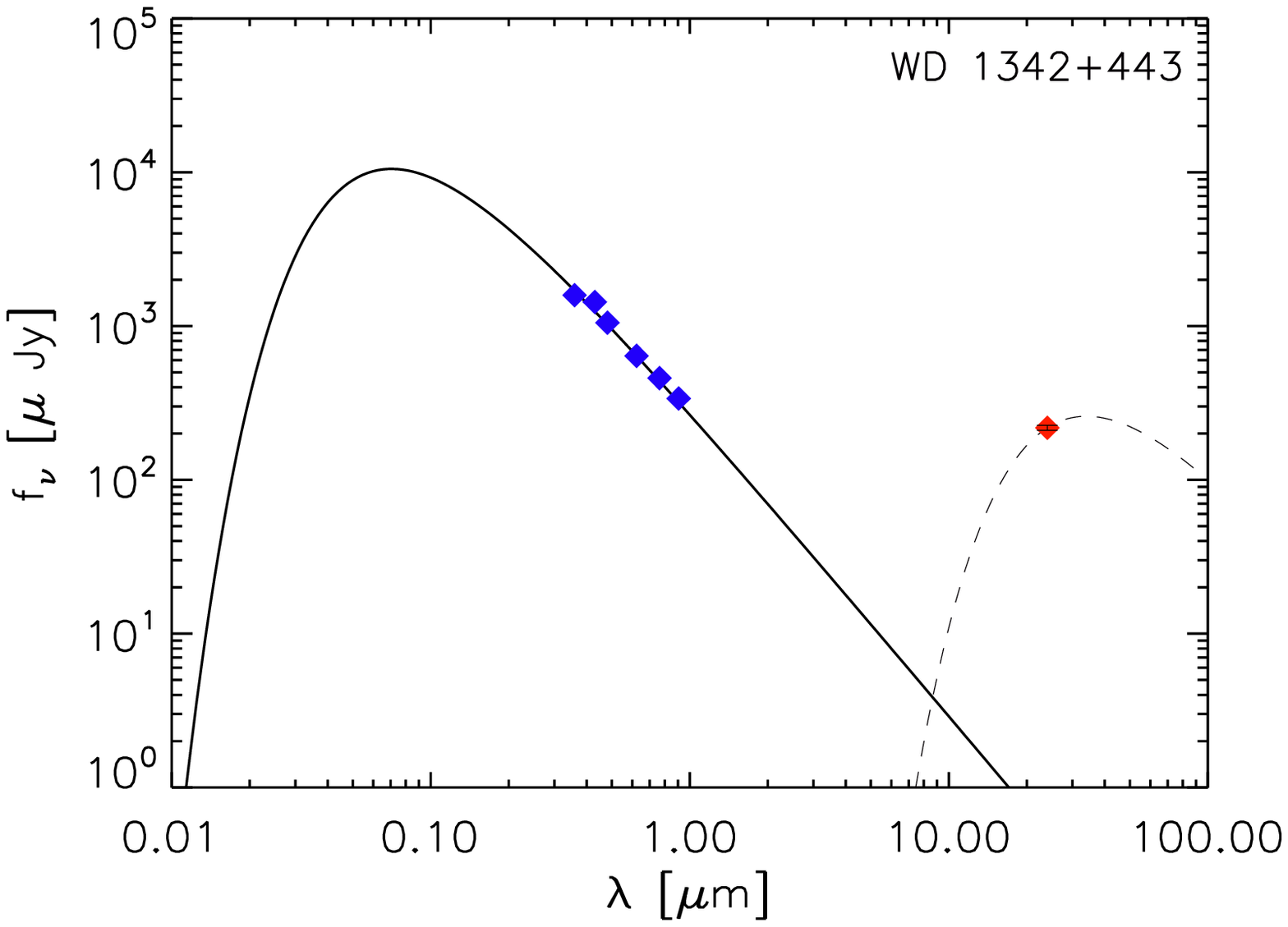}
\caption{Spectral energy distribution plots for WDs
detected in the MIPS 24 \um\ band. 
The photometric measurements include published optical magnitudes 
(blue diamonds), 2MASS $JHK$ (green diamonds), {\it Spitzer}
IRAC bands (yellow diamonds), and MIPS 24 and 70 \um\ bands 
(red diamonds). Solid blackbody lines in the optical wavelengths 
represent the WD photospheric emission, while dashed lines represent 
the blackbody-like excess emission with the best-fit dust 
temperatures (see Table 6).
For the four objects with excess emission in the IRAC bands, the
sum of the stellar emission and two blackbody components is plotted
in thin black line.}
\end{figure}

\clearpage
\begin{figure}  
\epsscale{0.7}
\plotone{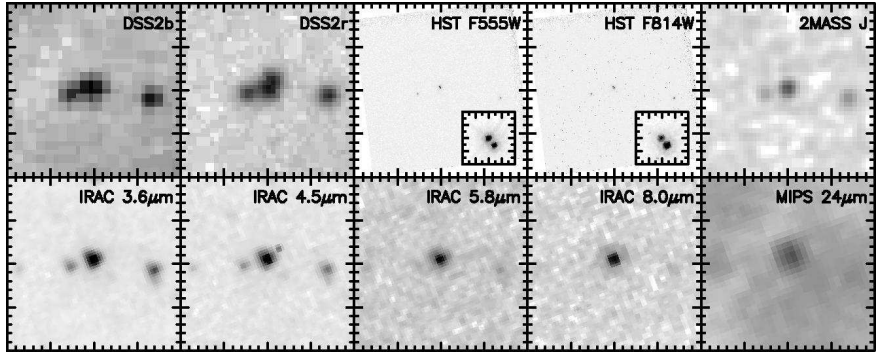}
\caption{Optical and IR images of CSPN K\,1-22.  The field of view of each panel 
is 40\arcsec$\times$40\arcsec.  The inset in the {\it HST} images
is 2\arcsec$\times$2\arcsec.}
\end{figure}

\begin{figure}  
\epsscale{0.85}
\plottwo{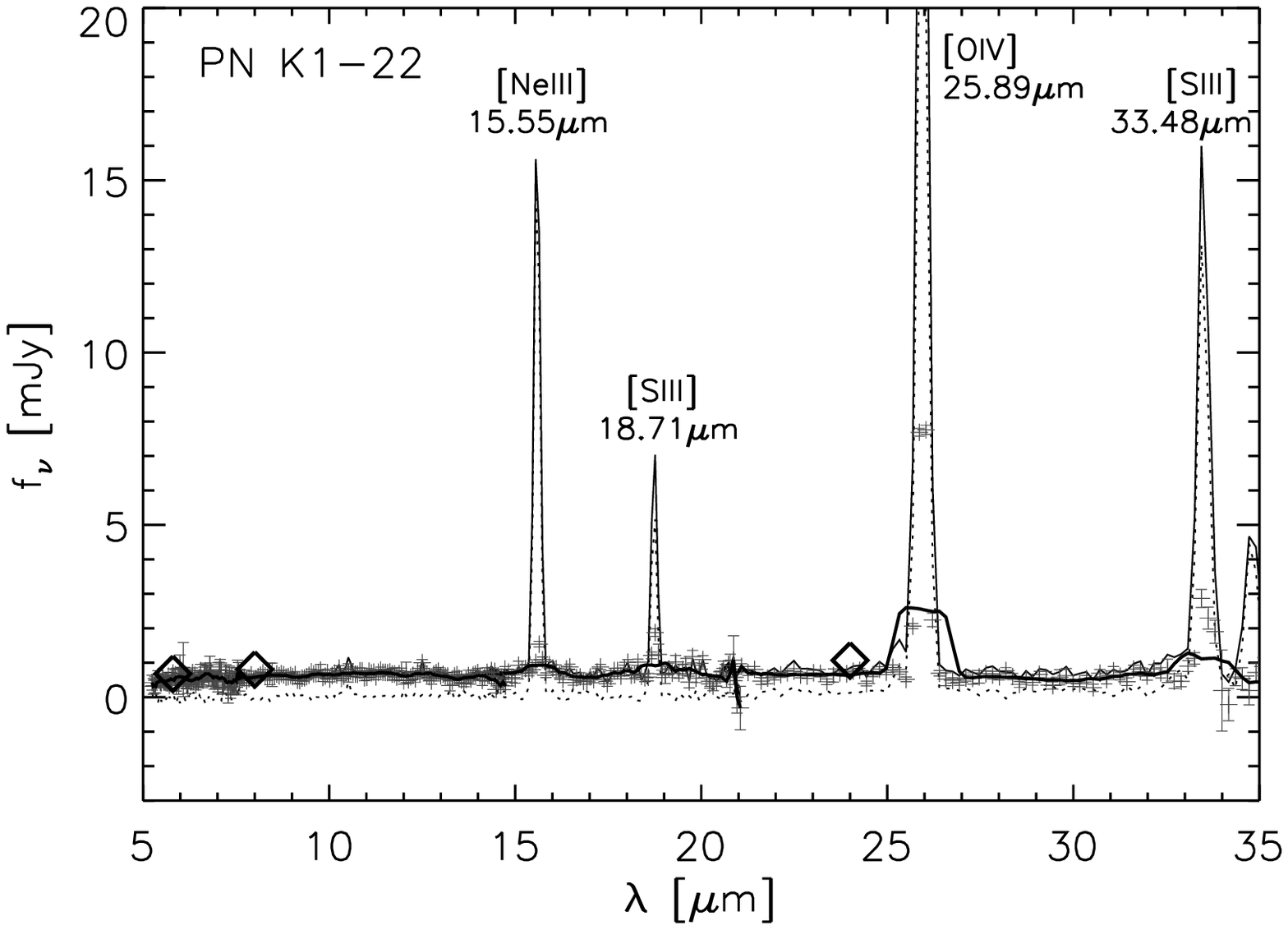}{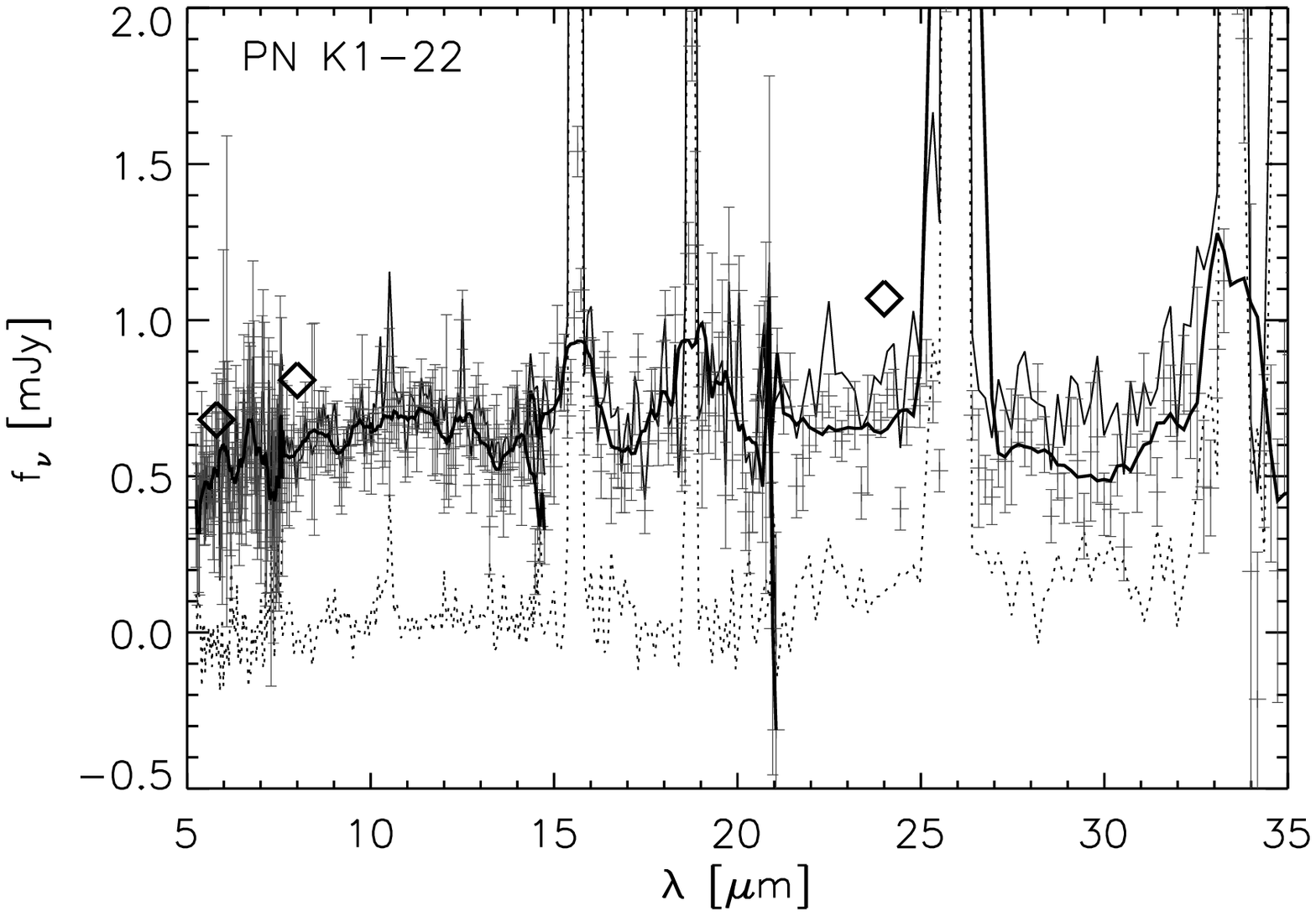}
\caption{{\it Spitzer} IRS spectrum of CSPN K\,1-22 plotted with different
stretches to illustrate the relative intensity of lines and continuum.
The spectrum extracted at CSPN K\,1-22 is plotted in a thin solid line,
the local background spectrum in a dotted line, the background-subtracted
spectrum of CSPN K\,1-22 in pixels with error bars, and the smoothed 
background-subtracted spectrum in a thick line.}
\end{figure}

\begin{figure}  
\epsscale{0.5}
\plotone{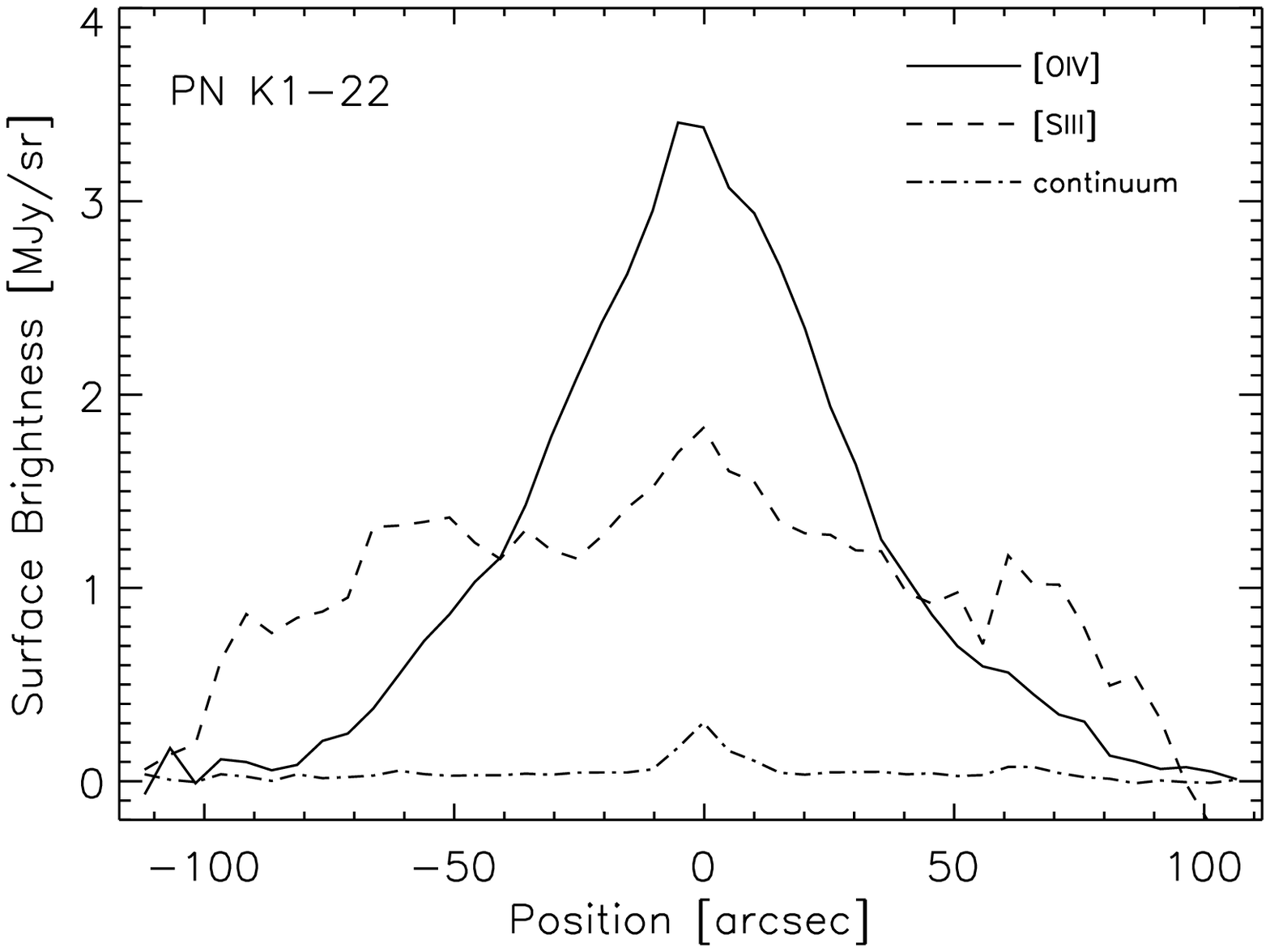}
\caption{Surface brightness profile plots of CSPN K\,1-22 extracted from
the {\it Spitzer } IRS data cube in the continuum (dash-dotted curve),
[\ion{O}{4}] 25.89 $\mu$m line (solid curve), and [\ion{S}{3}] 18.71 $\mu$m line
(dashed curve).}
\end{figure}

\newpage
\begin{figure}  
\epsscale{0.7}
\plotone{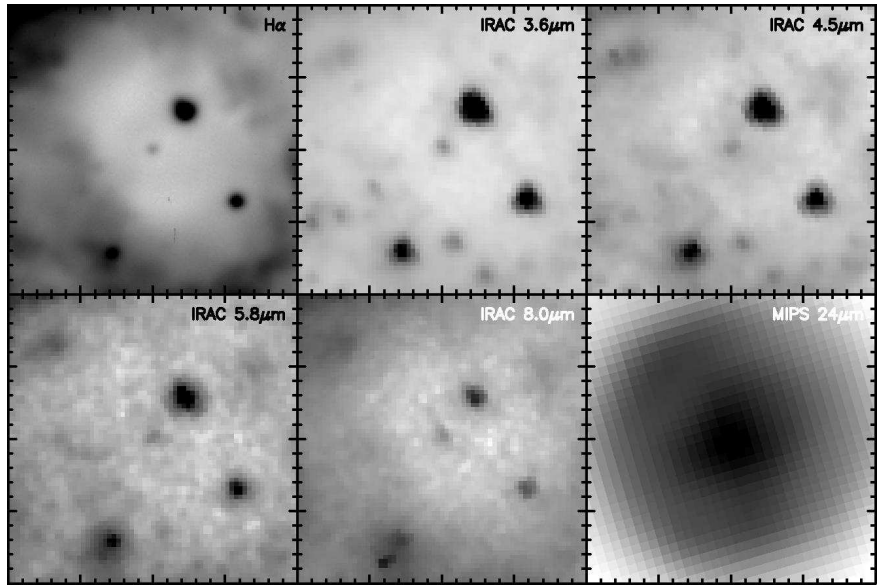}
\caption{H$\alpha$ and IR images of NGC\,2438.  The field of view of each panel 
is 40\arcsec$\times$40\arcsec.}
\end{figure}

\newpage
\begin{figure}  
\epsscale{0.7}
\plotone{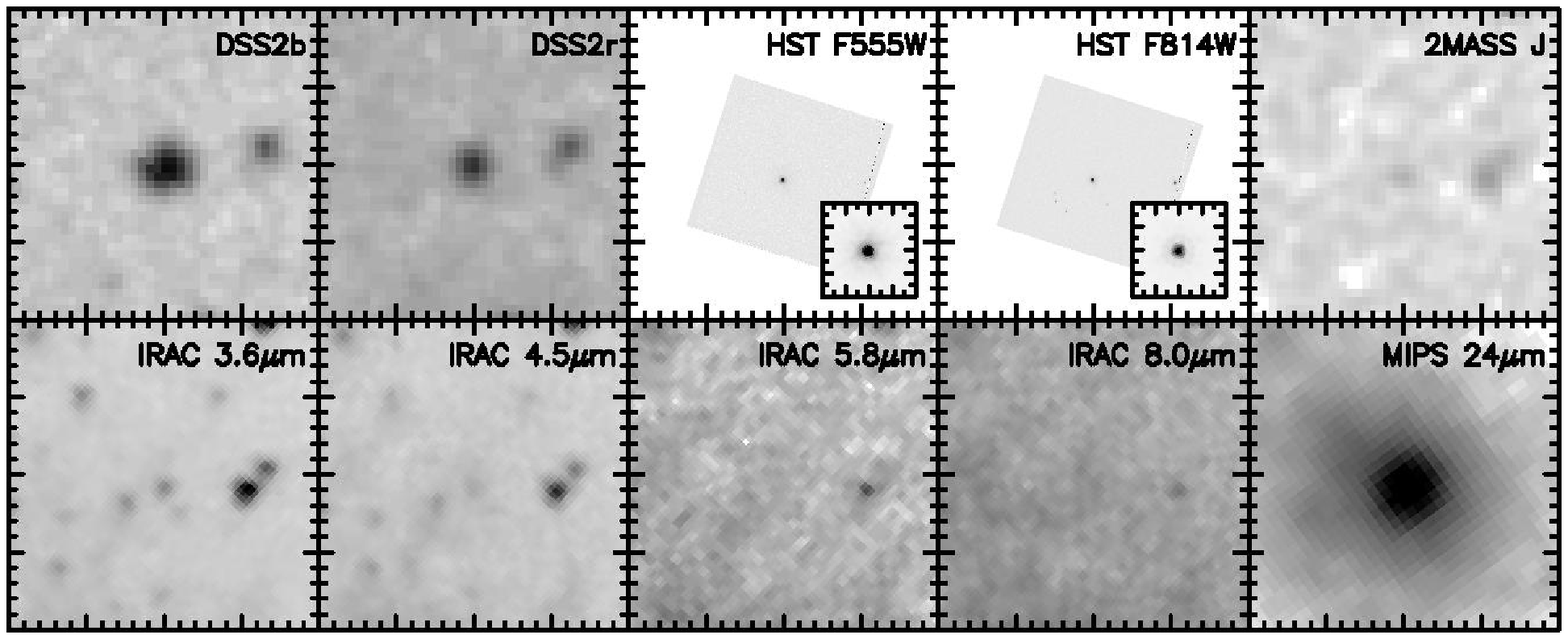}
\caption{Optical and IR mages of WD\,0103+732. The field of view of each panel 
is 40\arcsec$\times$40\arcsec.  The inset in the {\it HST} images
is 2\arcsec$\times$2\arcsec.}
\end{figure}

\newpage
\begin{figure}  
\epsscale{0.7}
\plotone{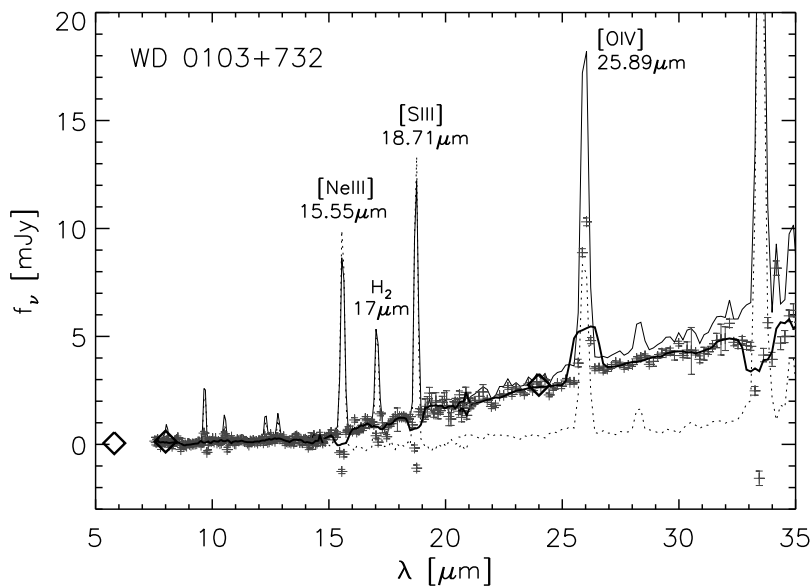}
\caption{{\it Spitzer} IRS spectrum of WD\,0103+732.  The spectrum 
extracted at WD\,0103+732 is plotted in a thin solid line,
the local background spectrum in a dotted line, the background-subtracted
spectrum of WD\,0103+732 in pixels with error bars, and the smoothed 
background-subtracted spectrum in a thick line.
}
\end{figure}

\begin{figure}  
\epsscale{0.7}
\plotone{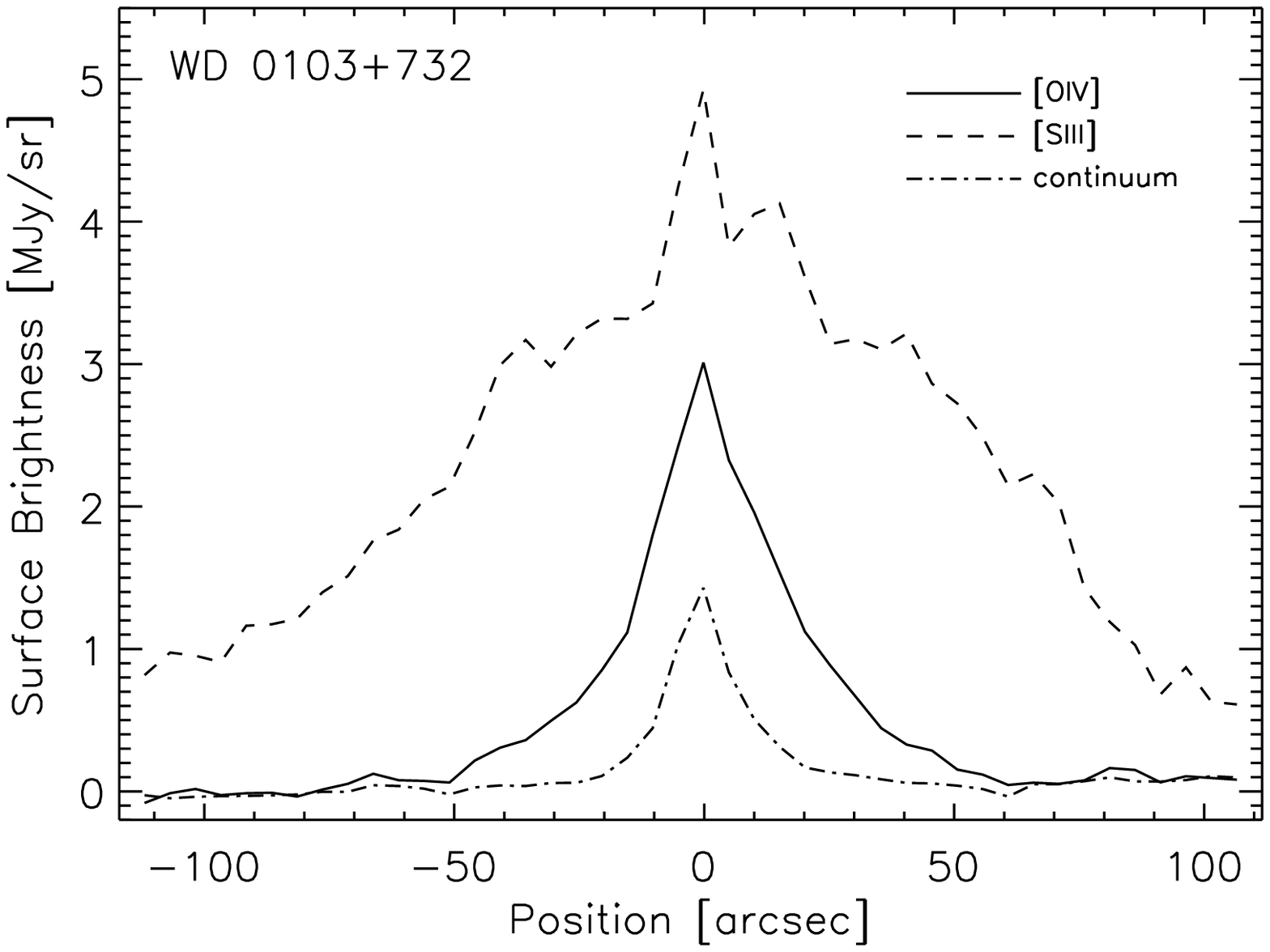}
\caption{Surface brightness profile plots of WD\,0103+732 extracted from
the {\it Spitzer } IRS data cube in the continuum (dash-dotted curve),
[\ion{O}{4}] 25.89 $\mu$m line (solid curve), and [\ion{S}{3}] 18.71 $\mu$m line
(dashed curve).}
\end{figure}

\newpage
\begin{figure}    
\epsscale{0.7}
\plotone{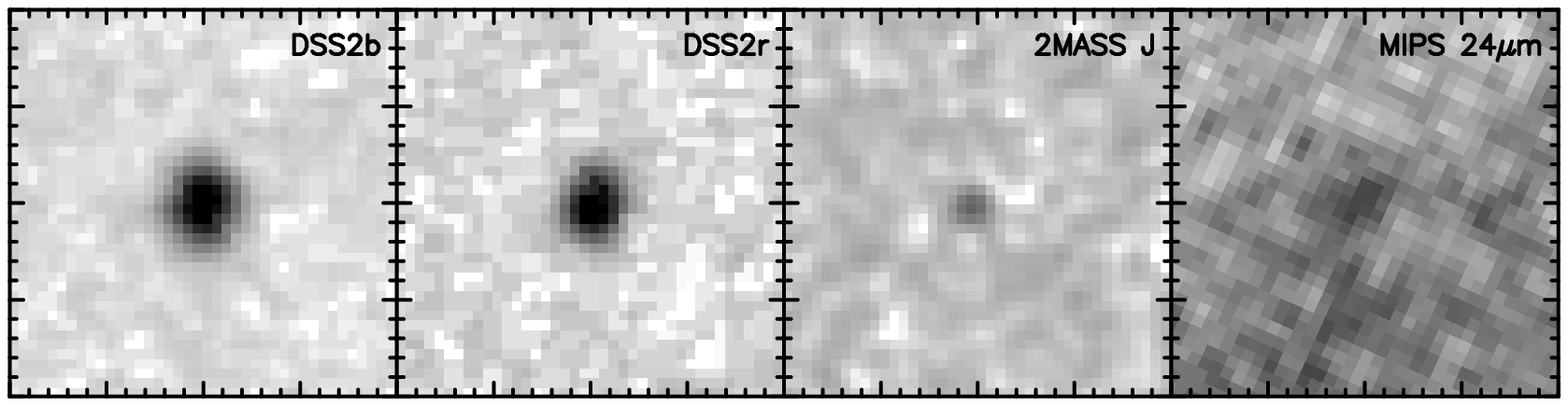}
\caption{Optical and IR images of WD\,0109+111.  The field of view of each panel 
is 40\arcsec$\times$40\arcsec.}
\end{figure}

\newpage
\begin{figure}     
\epsscale{0.7}
\plotone{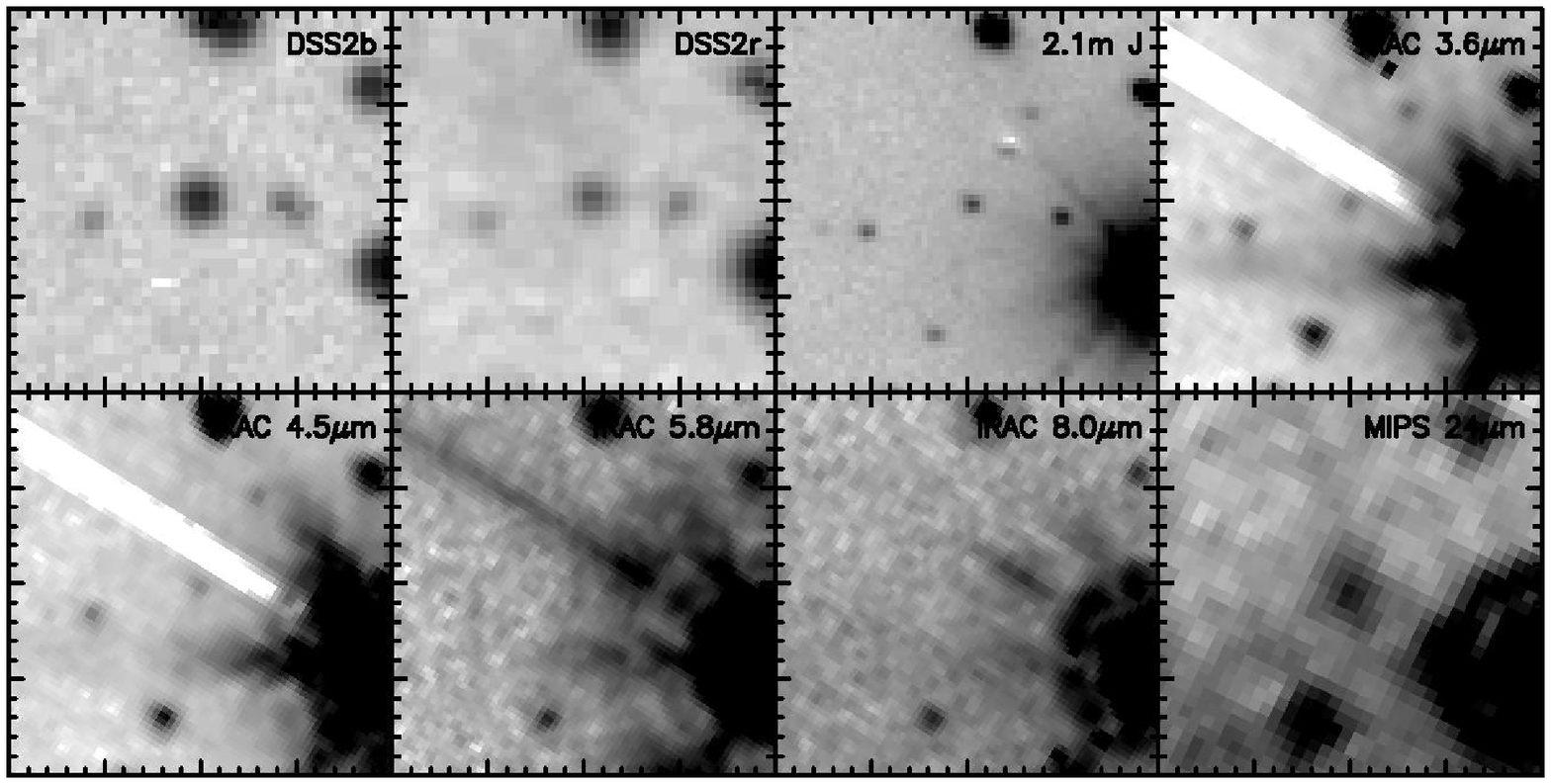}
\caption{Optical and IR images of WD\,0127+581.  The field of view of each panel 
is 40\arcsec$\times$40\arcsec.}
\end{figure}


\newpage
\begin{figure}     
\epsscale{0.7}
\plotone{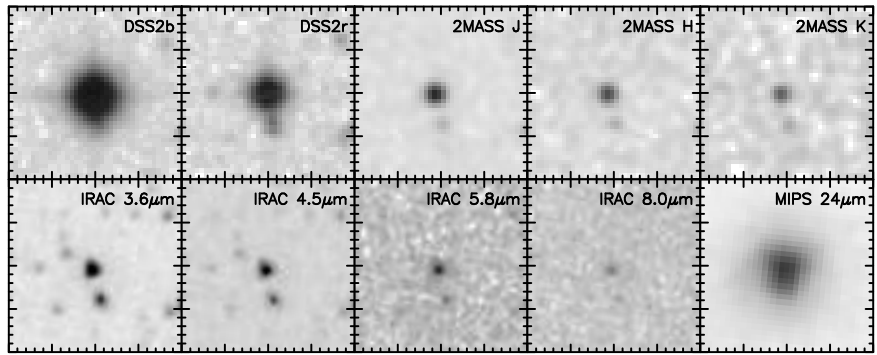}
\caption{Optical and IR images of WD\,0439+466.  The field of view of each panel 
is 40\arcsec$\times$40\arcsec.}
\end{figure}

\newpage
\begin{figure}     
\epsscale{0.75}
\plotone{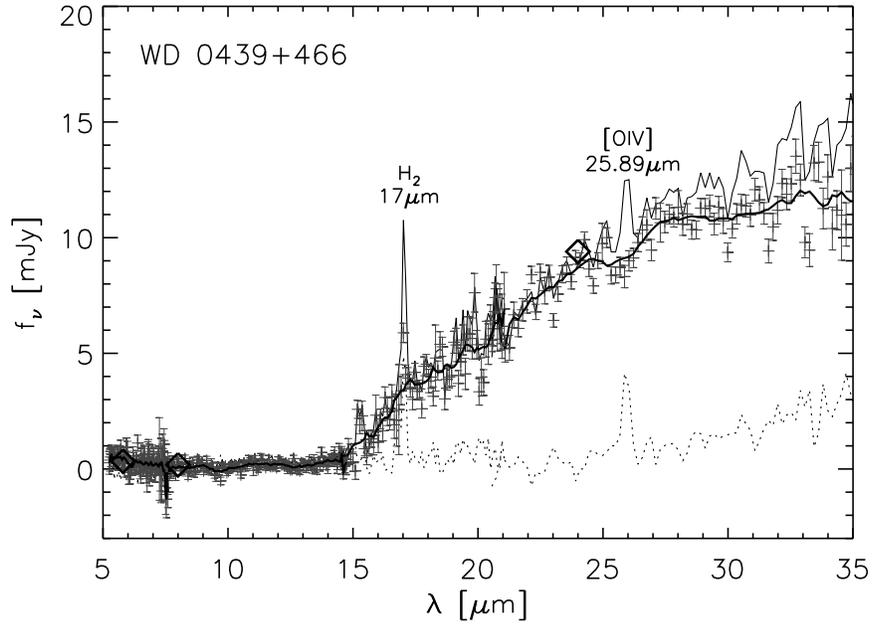}
\caption{{\it Spitzer} IRS spectrum of WD\,0439+466.  The spectrum 
extracted at WD\,0439+466 is plotted in a thin solid line,
the local background spectrum in a dotted line, the background-subtracted
spectrum of WD\,0439+466 in pixels with error bars, and the smoothed 
background-subtracted spectrum in a thick line.}
\end{figure}

\newpage
\begin{figure}     
\epsscale{0.7}
\plotone{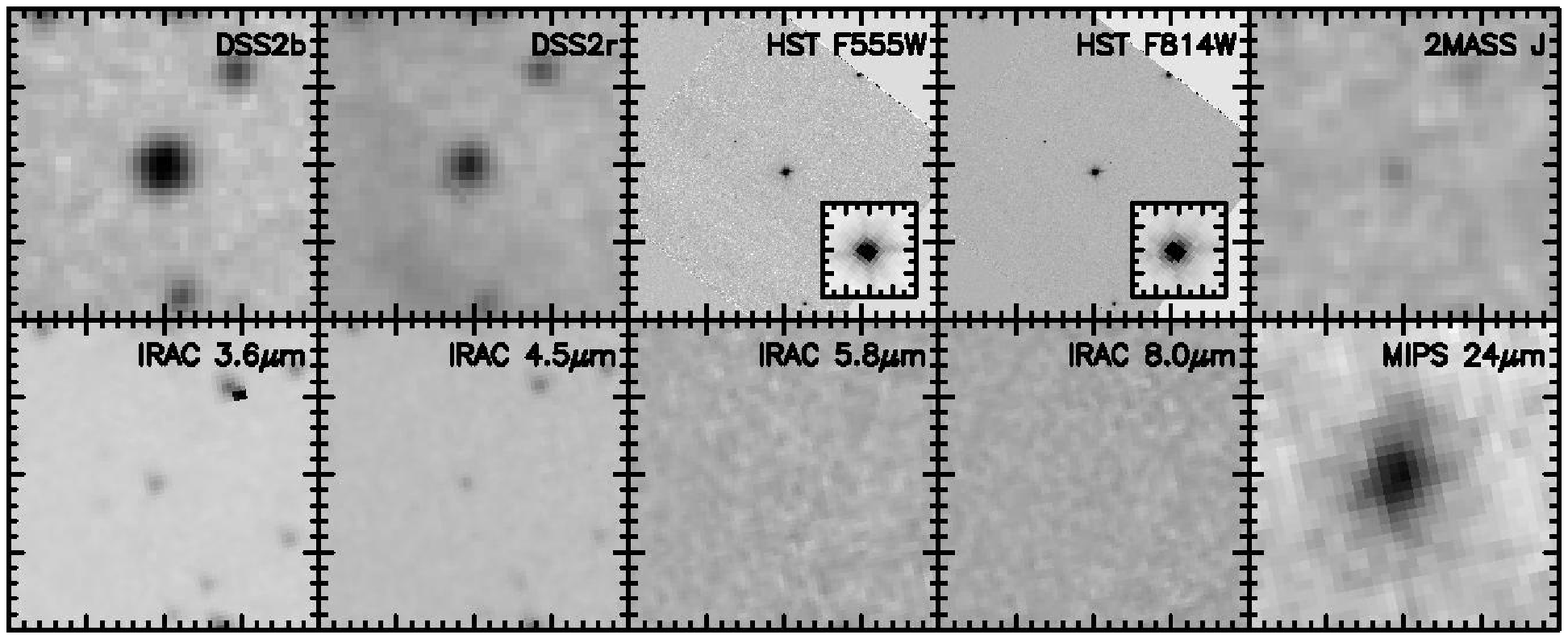}
\caption{Optical and IR images of WD\,0726+133.  The field of view of each panel 
is 40\arcsec$\times$40\arcsec.  The inset in the {\it HST} images
is 2\arcsec$\times$2\arcsec.}
\end{figure}

\newpage
\begin{figure}          
\epsscale{0.7}
\plotone{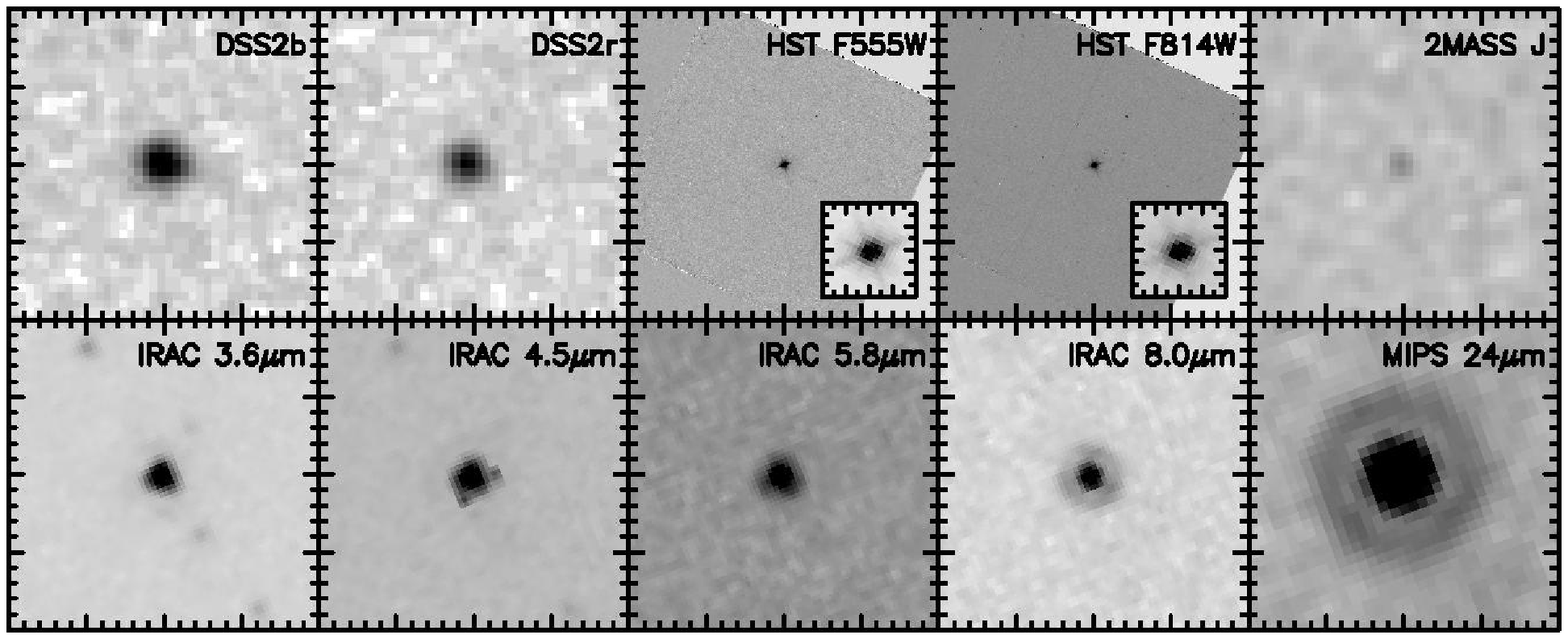}
\caption{Optical and IR images of WD\,0950+139.  The field of view of each panel 
is 40\arcsec$\times$40\arcsec.  The inset in the {\it HST} images
is 2\arcsec$\times$2\arcsec.}
\end{figure}

\newpage
\begin{figure}      
\epsscale{0.7}
\plotone{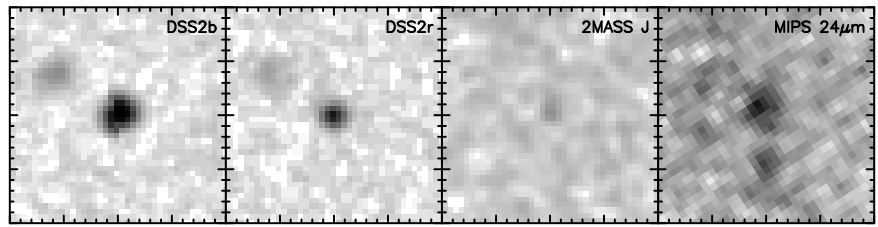}
\caption{Optical and IR images of WD\,1342+443.  The field of view of each panel 
is 40\arcsec$\times$40\arcsec.}
\end{figure}

\newpage
\begin{figure}      
\epsscale{0.8}
\plotone{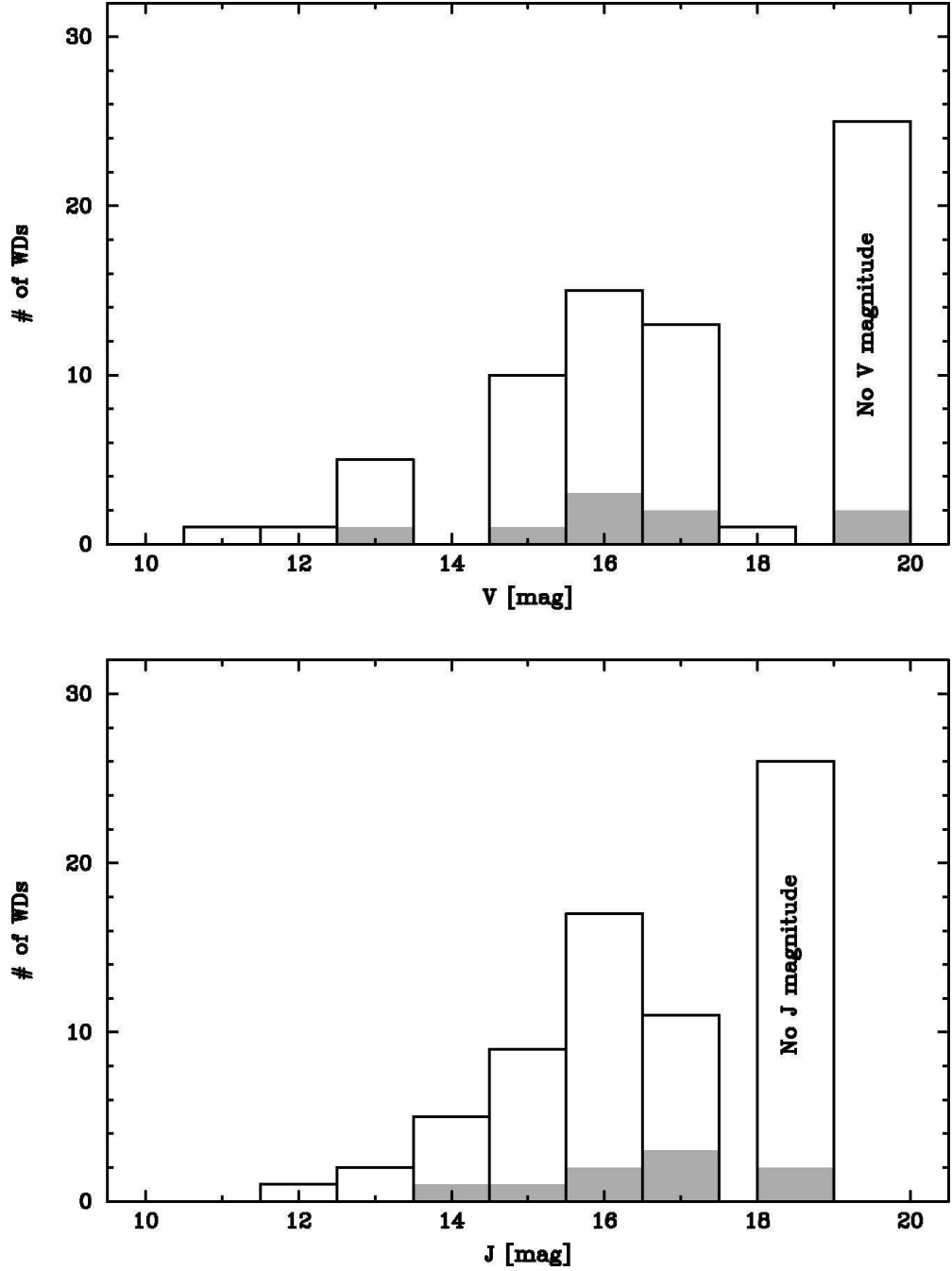}
\caption{Distribution of sample hot WDs in $J$ and $V$.  The WDs 
with 24 \um\ excesses are plotted in shaded boxes.}
\end{figure}

\newpage
\begin{figure}      
\epsscale{0.8}
\plotone{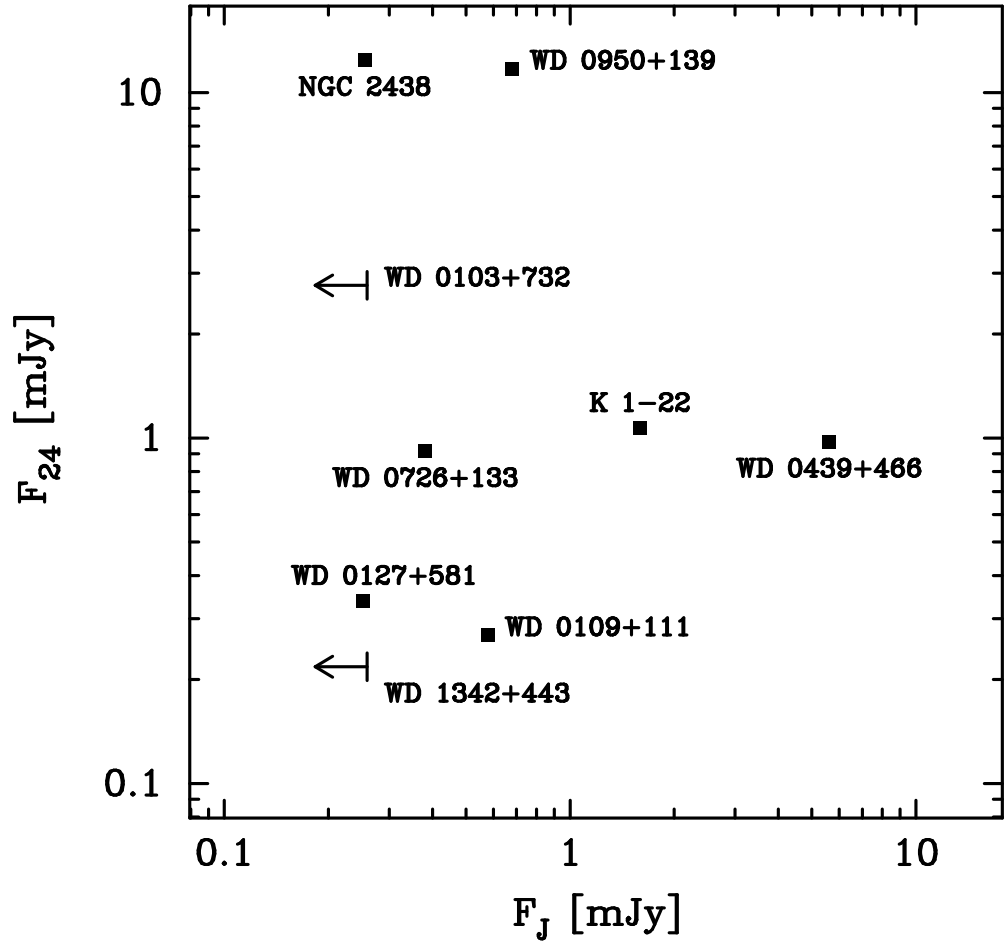}
\caption{Relationship between $F_J$ and $F_{24}$ 
for hot WDs showing 24 \um\ excesses.} 
\end{figure}
\clearpage

\begin{deluxetable}{llrllll}
\tablecolumns{4}
\tabletypesize{\scriptsize}
\tablewidth{0pc}
\tablecaption{Survey Sample of Hot White Dwarfs and Pre-White Dwarfs}
\tablehead{
~~~~~~~Star\tablenotemark{a} & ~~~~~~RA   & Dec~~~~~    & ~~~$T_{\rm eff}$ & ~~WD~  & ~~~PN~ & ~~~PNG \\
~~~~~~Name  & ~~~(J2000) & (J2000)~~~  &  (10$^3$ K)   & ~Type  & ~~Name & ~~Number 
}
\startdata
WD 0005+511 & 00 08 18.11 & +51 23 16.9 & ~~120 & DOQZ.4  &   ---     & ---    \\
PG 0038+199 & 00 41 35.31 & +20 09 17.5 & ~~115 & DO      &   ---     & ---  \\
WD 0044$-$121 & 00 47 03.31 &$-$11 52 18.9& ~~150 & PG1159  &  NGC 246  & 118.8$-$74.7 \\
WD 0103+732 & 01 07 07.74 & +73 33 25.2 & ~~150 & DA.34   &   EGB 1   & 124.0+10.7 \\
WD 0108+100 & 01 11 06.59 & +10 21 38.2 & ~~~84 & DOZ.6   &   ---     & ---    \\
WD 0109+111 & 01 12 23.06 & +11 23 36.1 & ~~110 & DOZ.46     &   ---     & --- \\
WD 0121$-$756 & 01 22 52.97 &$-$75 21 13.8& ~~180 & PG1159  &   ---     & --- \\
WD 0123$-$842 & 01 21 55.0  &$-$84 01 23.0& ~~--- & PG1159  &   ---     &	 --- \\
WD 0127+581 & 01 30 33.22 & +58 24 50.7 & ~~102 & DAO.49  &   Sh 2-188& 128.0$-$04.1 \\
WD 0130$-$196 & 01 32 39.31 &$-$19 21 40.4& ~~100 & PG1159.5&   ---     & --- \\
WD 0237+241 & 02 40 28.47 & +24 22 10.4 & ~~100 & DA0.5   &   ---     & --- \\
IC 289      & 03 10 19.36 & +61 19 00.5 & ~~100 & ---     &   IC 289  & 138.8+02.8 \\
WD 0316+002 & 03 18 58.28 & +00 23 25.8 & ~~100 & DA0.5   &   ---     & --- \\
WD 0322+452 & 03 27 15.42 & +45 24 20.2 & ~~125 & DAO.40  &   HDW 3   & 149.4$-$09.2 \\
NGC 1360    & 03 33 14.63 &$-$25 52 17.9& ~~110 & ---     &   NGC 1360& 220.3$-$53.9\\
WD 0439+466 & 04 43 21.20 & +46 42 06.4 & ~~~83 & DA.61   &   Sh 2-216& 158.5+00.7\\
WD 0444+049 & 04 47 04.51 & +04 58 41.7 & ~~100 & DQZO.5  &   ---     & ---   \\
WD 0500$-$156 & 05 03 07.51 &$-$15 36 22.5& ~~100 & DAO.51  &   Abell 7 & 215.5$-$30.8	\\
K 1-27      & 05 57 02.37 &$-$75 40 21.1& ~~105 & ---     &   K 1-27  & 286.8$-$29.5\\
WD 0556+106 & 05 59 24.87 & +10 41 39.9 & ~~141 & DA.36   &   WeDe 1  & 197.4$-$06.4 \\
WD 0615+556 & 06 19 33.95 & +55 36 43.7 & ~~~94 & DAO.54  &   PuWe 1  & 158.9+17.8\\
WD 0615+655 & 06 20 30.11 & +65 34 21.3 & ~~100 & DA.51   &   ---     & --- \\
Abell 15    & 06 27 02.04 &$-$25 22 49.3& ~~110 & ---     &   Abell 15& 233.5$-$16.3\\
Abell 20    & 07 22 57.64 & +01 45 33.4 & ~~119 & ---     &   Abell 20& 214.9+07.8\\
WD 0726+133 & 07 29 02.64 & +13 14 49.7 & ~~130\tablenotemark{b} & PG1159  &   Abell 21& 205.1+14.2\\
NGC 2438    & 07 41 50.50 &$-$14 44 07.7& ~~114 & ---     &   NGC 2438& 231.8+04.1 \\
WD 0753+535 & 07 57 51.69 & +53 25 17.5 & ~~125 & DQZO.4  &   JnEr 1  & 164.8+31.1 \\
WD 0823+316 & 08 27 05.57 & +31 30 08.2 & ~~100 & DA0.5   &   ---     & ---\\
NGC 2610    & 08 33 23.37 &$-$16 08 57.6& ~~100 & ---     &   NGC 2610& 239.6+13.9 \\
WD 0915+201 & 09 18 33.11 & +19 53 08.2 & ~~100 & DA0.5   &   ---     & ---\\
WD 0939+262 & 09 42 50.65 & +26 01 00.0 & ~~100 & DA0.5   &   ---     & ---\\
WD 0948+534 & 09 51 25.98 & +53 09 31.0 & ~~126 & DA.46   &   ---     & ---\\
LSS 1362    & 09 52 44.50 &$-$46 16 47.1& ~~100 & ---     &   HeDr 1  & 273.6+06.1\\
WD 0950+139 & 09 52 58.94 & +13 44 34.9 & ~~110 & DA.46   &   EGB 6   & 221.5+46.3\\
WD 1003$-$441 & 10 05 45.79 &$-$44 21 33.4& ~~120 & PG1159  &   Lo 4    & 274.3+09.1 \\
WD 1034+001 & 10 37 03.81 &$-$00 08 19.6& ~~100 & DOZ.5   &   ---     & ---\\
WD 1111+552 & 11 14 47.72 & +55 01 08.3 & ~~~94 & DAO.54  &   NGC 3587& 148.4+57.0\\
K 1-22      & 11 26 43.78 &$-$34 22 11.5& ~~141 & ---     &   K 1-22  & 283.6+25.3 \\
WD 1144+004 & 11 46 35.21 & +00 12 33.1 & ~~150 & DO.34   &   ---     & ---\\
LoTr 4      & 11 52 29.22 &$-$42 17 38.6& ~~120 & ---     &   LoTr 4  & 291.4+19.2 \\
BlDz 1      & 11 53 06.71 &$-$50 50 57.1& ~~128 & ---     &   BlDz 1  & 293.6+10.9 \\
BE UMa      & 11 57 44.84 & +48 56 17.9 & ~~105 & ---     &   BE UMa  & 144.8+65.8\\
WD 1159$-$034 & 12 01 45.97 &$-$03 45 41.3& ~~140 & DQZO.4  &   ---     & ---\\
WD 1253+378 & 12 55 14.77 & +37 32 29.7 & ~~100 & DA.5    &   ---     & ---\\
MeWe 1-3    & 13 28 04.90 &$-$54 41 58.4& ~~100 & ---     &   MeWe 1-3& 308.2+07.7 \\
WD 1342+443 & 13 44 26.87 & +44 08 33.3 & ~~~79 & DA.7    &   ---     & ---\\
WD 1424+534 & 14 25 55.40 & +53 15 25.2 & ~~110 & PG1159.46 &   ---     & ---\\
WD 1501+664 & 15 02 09.94 & +66 12 19.6 & ~~170 & DZQ.3     &   ---     & ---\\
WD 1517+740 & 15 16 46.23 & +73 52 07.0 & ~~110 & DO.5    &   ---     & ---\\
WD 1520+525 & 15 21 46.56 & +52 22 03.6 & ~~150 & PG1159.3 &   JavdSt 1& 085.4+52.3\\
WD 1522+662 & 15 22 56.70 & +66 04 41.5 & ~~140 & DO      &   ---     & ---\\
WD 1532+033 & 15 35 09.85 & +03 11 16.0 & ~~100 & DA.5    &   ---     & ---\\
WD 1547+015 & 15 49 44.98 & +01 25 54.6 & ~~100 & DA.5    &   ---     & ---\\
WD 1622+323 & 16 24 49.05 & +32 17 02.0 & ~~~78 & DA.65   &   ---     & ---\\
WD 1625+280 & 16 27 33.73 & +27 54 33.6 & ~~117 & DAO.43  &   Abell 39& 047.0+42.4\\
WD 1707+427 & 17 08 47.75 & +42 41 00.4 & ~~100 & DOZ.5   &   ---     & ---\\
WD 1729+583 & 17 29 50.37 & +58 18 09.0 & ~~~84 & DA0.6   &   ---     & ---\\
WD 1738+669 & 17 38 02.60 & +66 53 47.7 & ~~~95 & DA.53   &   ---     & ---\\
WD 1749+717 & 17 49 04.50 & +71 45 08.9 & ~~100 & DAO.5   &   ---     & ---	\\
WD 1751+106 & 17 53 32.27 & +10 37 23.7 & ~~117 & PG1159  &   Abell 43& 036.0+17.6\\
HaTr 7      & 17 54 09.30 &$-$60 49 57.1& ~~100 & ---     &   HaTr 7  & 332.5$-$16.9\\
WD 1827+778 & 18 25 08.72 & +77 55 37.1 & ~~~76 & DA.68    &   ---     & ---\\
WD 1830+721 & 18 30 04.90 & +72 11 34.5 & ~~100 & DO.5    &   ---     & ---\\
WD 1851$-$088 & 18 54 37.18 &$-$08 49 38.8& ~~~90 & DAO.56  &   IC 1295 & 025.4$-$04.7 \\
WD 1917+461 & 19 19 10.21 & +46 14 51.2 & ~~~88 & DAO.57  &   Abell 61& 077.6+14.7 \\
WD 1958+015 & 20 00 39.25 & +01 43 40.6 & ~~--- & PG1159  &   NGC 6852& 042.5$-$14.5\\
WD 2114+239 & 21 16 52.34 & +24 08 50.9 & ~~108 & DAO.47  &   Abell 74& 072.7$-$17.1 \\
WD 2115+339 & 21 17 08.30 & +34 12 27.3 & ~~170 & PG1159.3&   MWP 1   & 080.3$-$10.4\\
WD 2209+825 & 22 08 25.54 & +82 44 55.3 & ~~100 & DO      &   ---     & ---\\
WD 2246+066 & 22 49 25.29 & +06 56 45.8 & ~~100 & DA.51   &   ---     & ---\\
WD 2324+397 & 23 27 15.97 & +40 01 23.4 & ~~126 & DO.4    &   ---     & ---\\
WD 2333+301 & 23 35 53.33 & +30 28 06.2 & ~~170 & DOZ.3   &   Jn 1    & 104.2$-$29.6\\
\enddata
\tablenotetext{a}{For hot pre-WDs in PNe, the PN names are given.}
\tablenotetext{b}{\ion{He}{2} Zanstra temperature from \citet{Phillips03}.}
\end{deluxetable}

\begin{deluxetable}{llccrrcl}
\tabletypesize{\scriptsize}
\rotate
\tablewidth{0pc}
\tablecaption{MIPS 24 $\mu$m\ Measurements of Hot White Dwarfs and Pre-White Dwarfs}
\tablehead{
Star Name   & MIPS AOR &  PSF flux & Uncertainty & Upper Limit &Photometry & Photospheric  &   Remarks\\
                             &          &  (mJy) & 1-$\sigma$ (mJy) &3-$\sigma$ (mJy)& (mag)~~~~   &  (mJy)  & 
}
\startdata
      WD 0005+511  &  23366144  &      0.289 &      0.157 &       $<$ 0.761 &                        $K$ = 14.19  &   0.01164  &                                     \\
      PG 0038+199  &  23366400  &      0.014 &      0.050 &       $<$ 0.163 &                        $H$ = 15.37  &   0.00374  &                                     \\
    WD 0044$-$121  &  23366656  &     -2.677 &      9.641 &      $<$ 26.247 &                        $K$ = 12.87  &   0.03925  &       very bright nebular emission  \\
      WD 0103+732  &  23366912  &      2.760 &      0.141 &         \nodata &                        $V$ = 16.35  &   0.00058  &     superposed on diffuse emission  \\
      WD 0108+100  &  23367168  &      0.058 &      0.057 &       $<$ 0.230 &                        $J$ = 16.33  &   0.00130  &  a point source at 12\arcsec\ away  \\
      WD 0109+111  &  23367424  &      0.269 &      0.055 &         \nodata &                        $H$ = 16.05  &   0.00200  &                                     \\
    WD 0121$-$756  &  23367680  &      0.002 &      0.038 &       $<$ 0.116 &                        $H$ = 16.16  &   0.00181  &                                     \\
      WD 0127+581  &  23368192  &      0.338 &      0.142 &         \nodata &                        $V$ = 17.44  &   0.00021  &     superposed on diffuse emission  \\
    WD 0130$-$196  &  23368448  &      0.055 &      0.049 &       $<$ 0.201 &                        $K$ = 14.00  &   0.01386  &                                     \\
      WD 0237+241  &  23368704  &      0.073 &      0.067 &       $<$ 0.274 &                        $J$ = 16.28  &   0.00137  &    a point source at 9\farcs5 away  \\
      CSPN IC 289  &  23368960  &     1210.9 &      18.2  &      $<$ 1265.6 &                        $K$ = 15.14  &   0.00485  &       very bright diffuse emission  \\
      WD 0316+002  &  23369216  &     -0.026 &      0.056 &       $<$ 0.142 &                        $z$ = 19.66  &   0.00007  &                                     \\
      WD 0322+452  &  23369472  &      0.059 &      0.062 &       $<$ 0.245 &                        $V$ = 17.20  &   0.00026  &                                     \\
    CSPN NGC 1360  &  23369728  &     -0.605 &      1.752 &       $<$ 4.651 &                        $K$ = 12.37  &   0.06221  &            bright nebular emission  \\
      WD 0439+466  &  23369984  &      9.200 &      0.157\tablenotemark{a} &         \nodata &       $K$ = 13.66  &   0.01896  &     superposed on diffuse emission  \\
      WD 0444+049  &  23370240  &     -0.003 &      0.064 &       $<$ 0.187 &                            \nodata  &   \nodata  &  two sources at 10--11\arcsec\ away \\
    WD 0500$-$156  &  23370496  &      0.013 &      0.047 &       $<$ 0.156 &                        $H$ = 16.16  &   0.00181  &                                     \\
      CSPN K 1-27  &  23370752  &     11.603 &      0.372 &      $<$ 12.721 &                        $J$ = 16.40  &   0.00122  &            bright nebular emission  \\
      WD 0556+106  &  23371008  &      0.051 &      0.071 &       $<$ 0.264 &                        $V$ = 17.40  &   0.00022  &   a point source at 12\farcs5 away  \\
      WD 0615+556  &  23371264  &      0.110 &      0.049 &       $<$ 0.256 &                        $J$ = 15.90  &   0.00194  &   two sources at 3--4\arcsec\ away  \\
      WD 0615+655  &  23371520  &      0.028 &      0.042 &       $<$ 0.153 &                        $V$ = 15.70  &   0.00105  &                                     \\
    CSPN Abell 15  &  23371776  &     14.221 &      0.342 &      $<$ 15.248 &                        $V$ = 15.72  &   0.00103  &       very bright nebular emission  \\
    CSPN Abell 20  &  23372032  &      0.021 &      2.645 &       $<$ 7.956 &                        $V$ = 16.56  &   0.00048  &       very bright nebular emission  \\
      WD 0726+133  &  23372288  &      0.916 &      0.114 &         \nodata &                        $J$ = 16.58  &   0.00104  &     superposed on diffuse emission  \\
    CSPN NGC 2438  &  23372544  &     12.410 &     13.659 &         \nodata &                        $J$ = 17.02  &   0.00069  &     superposed on diffuse emission  \\
      WD 0753+535  &  23372800  &      0.197 &      0.159 &       $<$ 0.675 &                        $J$ = 16.58  &   0.00104  &       very bright nebular emission  \\
      WD 0823+316  &  23373056  &     -0.073 &      0.055 &       $<$ 0.092 &                        $K$ = 15.73  &   0.00282  &                                     \\
    CSPN NGC 2610  &  23373312  &      127.5 &        7.2 &       $<$ 149.0 &                        $H$ = 16.29  &   0.00160  &       very bright nebular emission  \\
      WD 0915+201  &  23373568  &      0.039 &      0.057 &       $<$ 0.210 &                        $V$ = 16.64  &   0.00044  &                                     \\
      WD 0939+262  &  23373824  &     -0.008 &      0.053 &       $<$ 0.151 &                        $H$ = 15.57  &   0.00311  &                                     \\
      WD 0948+534  &  23374080  &      0.060 &      0.043 &       $<$ 0.189 &                        $H$ = 16.15  &   0.00183  &                                     \\
    CSPN LSS 1362  &  23374336  &     -2.934 &      4.032 &       $<$ 9.164 &                        $K$ = 12.81  &   0.04148  &       very bright nebular emission  \\
      WD 0950+139  &  23374592  &     11.740 &      0.066 &         \nodata &                        $K$ = 16.10  &   0.00200  &                                     \\
    WD 1003$-$441  &  23374848  &      8.821 &      3.688 &      $<$ 19.887 &                        $V$ = 16.60  &   0.00046  &       very bright nebular emission  \\
      WD 1034+001  &  23375104  &      0.964 &      0.133 &       $<$ 1.364 &                        $K$ = 14.37  &   0.00986  &            bright nebular emission  \\
      WD 1111+552  &  23375360  &      2.190 &      1.320 &       $<$ 6.149 &                        $J$ = 16.71  &   0.00092  &            bright nebular emission  \\
      CSPN K 1-22  &  23375616  &      1.070 &      0.143 &         \nodata &                        $K$ = 14.27  &   0.01081  &     superposed on diffuse emission  \\
      WD 1144+004  &  23375872  &     -0.044 &      0.067 &       $<$ 0.158 &                        $H$ = 16.13  &   0.00186  &                                     \\
      CSPN LoTr 4  &  23376128  &     32.298 &      0.359 &      $<$ 33.375 &       $V$ = 16.50\tablenotemark{b}  &   0.00050  &       very bright nebular emission  \\
      CSPN BlDz 1  &  23376384  &      3.746 &      1.109 &       $<$ 7.073 &       $V$ = 18.40\tablenotemark{c}  &   0.00009  &            bright nebular emission  \\
      CSPN BE UMa  &  23376640  &      0.031 &      0.045 &       $<$ 0.165 &                        $K$ = 13.63  &   0.01949  &                                     \\
    WD 1159$-$034  &  23376896  &      0.044 &      0.058 &       $<$ 0.218 &                        $K$ = 15.73  &   0.00282  &                                     \\
      WD 1253+378  &  23377152  &      0.034 &      0.042 &       $<$ 0.159 &                        $K$ = 15.45  &   0.00365  &                                     \\
    CSPN MeWe 1-3  &  23377408  &     77.376 &      0.794 &      $<$ 79.759 &                             \nodata  &   \nodata  &       very bright nebular emission  \\
      WD 1342+443  &  23377664  &      0.218 &      0.041 &         \nodata &                        $z$ = 17.56  &   0.00049  &                                     \\
      WD 1424+534  &  23377920  &      0.141 &      0.044 &       $<$ 0.273 &                        $z$ = 16.97  &   0.00084  &      a point source 9\arcsec\ away  \\
      WD 1501+664  &  23378176  &      0.011 &      0.038 &       $<$ 0.125 &                        $V$ = 15.90  &   0.00087  &      a point source 2\arcsec\ away  \\
      WD 1517+740  &  23378432  &      0.005 &      0.038 &       $<$ 0.118 &                        $B$ = 15.90  &   0.00069  &                                     \\
      WD 1520+525  &  23378688  &      0.232 &      0.042 &       $<$ 0.358 &                        $J$ = 16.38  &   0.00125  &          in diffuse 24 um emission  \\
      WD 1522+662  &  23378944  &      0.016 &      0.036 &       $<$ 0.125 &                        $B$ = 16.40  &   0.00043  &                                     \\
      WD 1532+033  &  23379200  &      0.029 &      0.050 &       $<$ 0.179 &                        $H$ = 16.42  &   0.00142  &                                     \\
      WD 1547+015  &  23379456  &     -0.034 &      0.047 &       $<$ 0.107 &                        $J$ = 16.43  &   0.00119  &                                     \\
      WD 1622+323  &  23379712  &      0.039 &      0.038 &       $<$ 0.152 &                        $K$ = 13.77  &   0.01713  &                                     \\
      WD 1625+280  &  23379968  &      0.093 &      0.153 &       $<$ 0.553 &                        $J$ = 16.13  &   0.00157  &     superposed on diffuse emission  \\
      WD 1707+427  &  23380224  &     -0.069 &      0.038 &       $<$ 0.046 &                        $V$ = 16.70  &   0.00042  &                                     \\
      WD 1729+583  &  23380480  &     -0.025 &      0.038 &       $<$ 0.090 &                        $z$ = 19.76  &   0.00007  &                                     \\
      WD 1738+669  &  23380736  &      0.021 &      0.036 &       $<$ 0.129 &                        $K$ = 15.45  &   0.00365  &      a point source 12\farcs5 away  \\
      WD 1749+717  &  23380992  &      0.004 &      0.039 &       $<$ 0.123 &                        $J$ = 16.78  &   0.00086  &                                     \\
      WD 1751+106  &  23381248  &     -1.840 &      2.641 &       $<$ 6.084 &                        $K$ = 15.33  &   0.00407  &            bright nebular emission  \\
      CSPN HaTr 7  &  23381504  &      0.194 &      0.294 &       $<$ 1.075 &                        $K$ = 15.75  &   0.00277  &            bright nebular emission  \\
      WD 1827+778  &  23381760  &      0.022 &      0.037 &       $<$ 0.133 &                        $J$ = 16.64  &   0.00098  &      a point source 12\farcs5 away  \\
      WD 1830+721  &  23382016  &     -0.006 &      0.040 &       $<$ 0.113 &                        $B$ = 17.00  &   0.00025  &      a point source 7\arcsec\ away  \\
    WD 1851$-$088  &  23382272  &      5.071 &      5.496 &      $<$ 21.557 &       $V$ = 16.90\tablenotemark{d}  &   0.00035  &       very bright nebular emission  \\
      WD 1917+461  &  23382528  &      0.049 &      0.090 &       $<$ 0.318 &                        $V$ = 17.39  &   0.00022  &            bright nebular emission  \\
      WD 1958+015  &  23382784  &    136.0 &      1.5 &     $<$ 140.4 &                        $V$ = 17.90  &   0.00014  &       very bright nebular emission  \\
      WD 2114+239  &  23383040  &     -0.073 &      0.061 &       $<$ 0.109 &                        $V$ = 17.05  &   0.00030  &                                     \\
      WD 2115+339  &  23383296  &     -0.120 &      0.341 &       $<$ 0.903 &                        $K$ = 14.18  &   0.01174  &            bright nebular emission  \\
      WD 2209+825  &  23383552  &     -0.010 &      0.040 &       $<$ 0.110 &                        $J$ = 16.58  &   0.00104  &                                     \\
      WD 2246+066  &  23383808  &      0.067 &      0.052 &       $<$ 0.222 &       $B$ = 16.80\tablenotemark{e}  &   0.00030  &                                     \\
      WD 2324+397  &  23384064  &      0.003 &      0.075 &       $<$ 0.227 &                        $K$ = 15.40  &   0.00382  &   near a patch of diffuse emission  \\
      WD 2333+301  &  23384320  &      0.363 &      0.154 &       $<$ 0.824 &                        $J$ = 16.70  &   0.00093  &  complex emission, no point source \\
\enddata
\tablenotetext{a}{Aperture photometry flux.}
\tablenotetext{b}{From \citet{Retal96}.}
\tablenotetext{c}{From \citet{Retal99}.}
\tablenotetext{d}{From \citet{NS95}.}
\tablenotetext{e}{From \citet{Hetal98}.}

\end{deluxetable}

\begin{deluxetable}{lcccccc}
\rotate
\tablewidth{0pc}
\tablecaption{{\it Spitzer} Photometry of Hot White Dwarfs with 24 \um\ Excesses}
\tablehead{
              &  $f_{3.6 \mu {\rm m}}$  &   $f_{4.5 \mu {\rm m}}$  &   $f_{5.8 \mu {\rm m}}$   &   $f_{8.0 \mu {\rm m}}$   
    &  $f_{24 \mu {\rm m}}$ & $f_{70 \mu {\rm m}}$ \\
WD Name       &  ($\mu$Jy)      &     ($\mu$Jy)    &    ($\mu$Jy)      &    ($\mu$Jy)      &    ($\mu$Jy) & ($\mu$Jy) }
\startdata
CSPN K\,1-22  & 829$\pm$42  & 706$\pm$36   &  681$\pm$38   &   808$\pm$42  &  1,070$\pm$143    & $<$12,000 \\
CSPN NGC\,2438& 103$\pm$40  &  91$\pm$57   &   82$\pm$58   &   117$\pm$95  & 12,410$\pm$13,700\tablenotemark{a} & ... \\
WD\,0103+732  &  88$\pm$15  &  71$\pm$19   &   58$\pm$20   &   132$\pm$35  &  2,760$\pm$141   & $<$55,000\\
WD\,0109+111  &    ...          &      ...         &      ...          &      ...          &    269$\pm$55   & ... \\
WD\,0127+581  &  76$\pm$47  &  56$\pm$26   &  185$\pm$161  &    92$\pm$45  &    338$\pm$142     & ... \\
WD\,0439+466  & 862$\pm$44  & 514$\pm$26   &  336$\pm$22   &   180$\pm$14  &  9,200$\pm$157    & 9,200 $\pm$8,400\\
WD\,0726+133  & 37.7$\pm$2.5  &  22.8$\pm$2.0    &   $<$19.8         &   $<$17.0         &    916$\pm$114   &  ... \\
WD\,0950+139  & 977$\pm$15\tablenotemark{b}&  1176$\pm$15\tablenotemark{b}& 1773$\pm$36\tablenotemark{b}&    3772$\pm$37\tablenotemark{b}& 11,740$\pm$66   &  ... \\
WD\,1342+443  &    ...          &      ...         &      ...          &      ...          &    218$\pm$41     &  ...\\
\enddata
\tablenotetext{a}{The photometric uncertainty is dominated by the
  bright nebular emission.}
\tablenotetext{b}{Su et al.\ 2011, in preparation.}
\end{deluxetable}

\begin{deluxetable}{llccccc}
\tablewidth{0pc}
\tablecaption{Summary of the IRS Follow-up Observations\label{tab_IRSsetup}}
\tablehead{
                 & \colhead{Date}     & \colhead{IRS}  & \colhead{SL1} & \colhead{SL2}  & \colhead{LL1} & \colhead{LL2} \\
\colhead{Star Name}& \colhead{Observed} & \colhead{Mode} &  5.2--8.7 \um  & 7.4--14.5 \um & 14.0--21.3 \um & 19.5--38.0 \um 
}
\startdata
CSPN K1-22     & 2009 Feb 26 & staring & 7$\times$60~s &  7$\times$60~s & 8$\times$120~s & 8$\times$120~s \\
WD\,0103+732   & 2008 Sep 13 & staring & 8$\times$60~s &  8$\times$60~s & 4$\times$120~s & 8$\times$120~s \\
WD\,0127+581   & 2008 Oct 09 & mapping\tablenotemark{a} & 2$\times$60~s &  2$\times$60~s & 2$\times$120~s & 2$\times$120~s \\
WD\,0439+466   & 2008 Sep 7 & mapping\tablenotemark{a} & 8$\times$60~s & 16$\times$60~s & 4$\times$120~s & 4$\times$120~s 
\enddata
\tablenotetext{a}{Exposure times given are for each pointing in the map.}
\end{deluxetable}

\clearpage
\begin{deluxetable}{lcccc}
\tablecaption{Source and Background Apertures for IRS Spectral Extractions}
\tablehead{
\colhead{WD} &
\colhead{LL1 \& LL2} &
\colhead{SL1 \& SL2} &
\colhead{LL1 \& LL2} &
\colhead{SL1 \& SL2} \\
\colhead{Name} &
\colhead{Aperture (arcsec$^2$)} &
\colhead{Aperture (arcsec$^2$)} &
\colhead{Offset (arcsec)} &
\colhead{Offset (arcsec)}
}
\startdata
CSPN K1-22    & 156 & 35 & 20 & 9.3\\
WD 0103+732   & 155 & 41 & 20 & 13.0\\
WD 0439+466   & 413 & 96 & 25.4 & 16.7\\
\enddata
\label{SED_values}
\end{deluxetable}

\newpage

\begin{deluxetable}{llrlrrclll}
\tabletypesize{\scriptsize}
\rotate
\tablewidth{0pc}
\tablecaption{$L_{\rm IR}/L_*$ of Hot White Dwarfs with 24 \um\ Excesses}
\tablehead{
WD                     &  Spec  & $T_{\rm eff}$~~~  &          & Distance & $R_*$~   & $T_{\rm dust}$  &      & SED  &   \\
Name                          &  Type  & (K)~~~    & $E(B-V)$ &  (pc)~~~    & ($R_\oplus$) &(K)     & $L_{\rm IR}/L_*$~    & Type\tablenotemark{a}
& References\tablenotemark{b}}
\startdata
CSPN K\,1-22                  & CSPN   &  141,000 & ~~~0.076    & 1,330~~~    & 3.3~~      & 700+150  &  1.1$\times 10^{-4}$ & EGB\,6-like & 1, 2 \\
CSPN NGC\,2438                & CSPN   &  114,000 & ~~~0.25     & 1,200~~~    & 4.1~~      & 1200+150 &  4.7$\times 10^{-4}$ & EGB\,6-like & 1, 3 \\
WD\,0103+732 (CSPN EGB\,1)    & DA.34  &  150,000 & ~~~0.58     &  650~~~     & 3.7~~      & 190~     &  1.4$\times 10^{-5}$ & Helix-like  & 4, 5 \\
WD\,0109+111                  & DOZ.46 &  110,000 & ~~~0.065    &  280~~~     & 2.3~~      & 150~     &  4.9$\times 10^{-6}$ & Helix-like  & 6, 7, 8\\
WD\,0127+581 (CSPN Sh\,2-188) & DAO.49 &  102,000 & ~~~0.27     &  600~~~     & 1.7~~      & 900+150  &  6.6$\times 10^{-5}$ & EGB\,6-like & 4, 9\\
WD\,0439+466 (CSPN Sh\,2-216) & DA.61  &  83,000  & ~~~0.065    &  129~~~     & 2.4~~      & 150~     &  2.4$\times 10^{-5}$ & Helix-like  & 10, 11\\
WD\,0726+133 (CSPN Abell\,21) & PG1159 &  130,000 & ~~~0.13     &  541~~~     & 2.0~~      & 150~     &  1.6$\times 10^{-5}$ & Helix-like  & 2, 11, 12\\
WD\,0950+139 (CSPN EGB\,6)    & DA.46  &  110,000 & ~~~0.21     &  645~~~     & 3.0~~      & 500+150  &  4.7$\times 10^{-4}$ & EGB\,6-like & 5, 13\\
WD\,1342+443                  & DA.7   &  79,000  & ~~~---      &  437~~~     & 1.4~~      & 150~     &  5.1$\times 10^{-5}$ & Helix-like  & 13 \\
WD\,2226$-$210 (CSPN Helix)   & DAO.49 &  110,000 & ~~~0.03     &  210~~~     & 2.6~~      & 120~     &  2.5$\times 10^{-4}$ & Helix-like  & 11, 14\\
\enddata
\tablenotetext{a}{EGB\,6-like SEDs show excesses in the IRAC bands as well as the MIPS
24 \um\ band.  Helix-like SEDs show no excess emission at wavelengths shorter than $\sim$8 \um.}
\tablenotetext{b}{References for stellar temperature, extinction, and distance: (1) \citet{Retal99}, (2) \citet{Cetal99}, (3) \citet{Ph04},
 (4) \citet{Nap01}, (5) \citet{Tetal92}, (6) \citet{DW96}, (7) \citet{Wetal97}, (8) \citet{Wetal85}, (9) \citet{KJ89},
 (10) \citet{Retal09}, (11) \citet{Hetal07}, (12) \citet{Phillips03}, (13) \citet{Letal05}, (14) \citet{Nap99}.}
\end{deluxetable}


\begin{thebibliography}{}

\bibitem[Becklin et al.(2005)]{Betal05} 
Becklin, E.~E., Farihi, J., Jura, M., Song, I., Weinberger, A.~J., 
\& Zuckerman, B.\ 2005, \apjl, 632, L119 

\bibitem[Bilikova et al.(2011a)]{Betal11a} 
Bilikova, J., Chu, Y.-H., Gruendl, R., Su, K., 
\& Rauch, T.\ 2011a, Asymmetric Planetary Nebulae 5 Conference


\bibitem[Bilikova et al.(2009)]{Betal09} 
Bilikova, J., Chu, Y.-H., Su, K., Gruendl, R., Rauch, T., De Marco, O., 
\& Volk, K.\ 2009, Journal of Physics Conference Series, 172, 012055

\bibitem[Bond(1994)]{Bond94} 
Bond, H.~E.\ 1994, Interacting Binary Stars, Asp.\ Conf.\ Series, 56, 179

\bibitem[Bond(2009)]{Bond09} 
Bond, H.~E.\ 2009, Journal of Physics Conference Series, 172, 012029 

\bibitem[Bonsor \& Wyatt(2010)]{BW10} 
Bonsor, A., \& Wyatt, M.\ 2010, \mnras, 409, 1631

\bibitem[Carpenter et al.(2009)]{Caetal09} 
Carpenter, J.~M., et al.\ 2009, \apjs, 181, 197

\bibitem[Chu et al.(2009)]{Cetal09}
Chu, Y.-H., et al.\ 2009, \aj, 138, 691

\bibitem[Chu et al.(2011)]{Cetal11} 
Chu, Y.-H., Gruendl, R.~A., Bil{\'{\i}}kov{\'a}, J., Riddle, A., 
\& Su, K.~Y.-L.\ 2011, American Institute of Physics Conference 
Series, 1331, 230

\bibitem[Ciardullo et al.(1999)]{Cetal99} 
Ciardullo, R., Bond, H.~E., Sipior, M.~S., Fullton, L.~K., Zhang, 
C.-Y., \& Schaefer, K.~G.\ 1999, \aj, 118, 488

\bibitem[de Marco(2009)]{dM09} 
de Marco, O.\ 2009, \pasp, 121, 316 

\bibitem[de Ruyter et al.(2006)]{DRetal06} 
de Ruyter, S., van Winckel, H., Maas, T., Lloyd Evans, T., Waters, L.~B.~F.~M., 
\& Dejonghe, H.\ 2006, \aap, 448, 641 

\bibitem[Debes et al.(2011)]{Detal11}
Debes, J.\ H., et al.\ 2011, \apj, in press (arXiv:1012.4859v1)

\bibitem[Debes \& Sigurdsson(2002)]{DS02} 
Debes, J.~H., \& Sigurdsson, S.\ 2002, \apj, 572, 556 

\bibitem[Diolaiti et al.(2000)]{diolaiti00} 
Diolaiti, E., Bendinelli, O., Bonaccini, D., Close, L., Currie, D., 
\& Parmeggiani, G.\ 2000, \aaps, 147, 335 

\bibitem[Dole et al.(2004)]{dole04} 
Dole, H., et al.\ 2004, \apjs, 154, 93 

\bibitem[Dong et al.(2010)]{Detal10} 
Dong, R., Wang, Y., Lin, D.~N.~C., \& Liu, X.-W.\ 2010, \apj, 715, 1036

\bibitem[Dreizler \& Werner(1996)]{DW96} 
Dreizler, S., \& Werner, K.\ 1996, \aap, 314, 217

\bibitem[Eisenstein et al.(2006)]{Eetal06} 
Eisenstein, D.~J., et al.\ 2006, \apjs, 167, 40

\bibitem[Engelbracht et al.(2007)]{Enetal07} 
Engelbracht, C.~W., et al.\ 2007, \pasp, 119, 994

\bibitem[Farihi et al.(2010)]{Fetal10}
Farihi, J., Jura, M., Lee, J.-E., \& Zuckerman, B.\ 2010, \apj, 714, 1386

\bibitem[Fazio et al.(2004)]{fazio04} 
Fazio, G.~G., et al.\ 2004, \apjs, 154, 10 

\bibitem[Fulbright \& Liebert(1993)]{FL93} 
Fulbright, M.~S., \& Liebert, J.\ 1993, \apj, 410, 275

\bibitem[Gordon et al.(2005)]{Getal05} 
Gordon, K.~D., et al.\ 2005, \pasp, 117, 503

\bibitem[Gordon et al.(2007)]{Getal07} 
Gordon, K.~D., et al.\ 2007, \pasp, 119, 1019 

\bibitem[Harris et al.(2007)]{Hetal07} 
Harris, H.~C., et al.\ 2007, \aj, 133, 631

\bibitem[Homeier et al.(1998)]{Hetal98} 
Homeier, D., et al.\ 1998, \aap, 338, 563

\bibitem[Houck et al.(2004)]{Hetal04} 
Houck, J.~R., et al.\ 2004, \apjs, 154, 18

\bibitem[Jura(2003)]{Jura03} 
Jura, M.\ 2003, \apjl, 584, L91

\bibitem[Jura et al.(2007)]{Jetal07} 
Jura, M., Farihi, J., \& Zuckerman, B.\ 2007, \apj, 663, 1285

\bibitem[Kilic \& Redfield(2007)]{KR07} 
Kilic, M., \& Redfield, S.\ 2007, \apj, 660, 641

\bibitem[Kilic et al.(2005)]{Ketal05} 
Kilic, M., von Hippel, T., Leggett, S.~K., \& Winget, D.~E.\ 2005,
\apjl, 632, L115

\bibitem[Kilic et al.(2006)]{Ketal06} 
Kilic, M., von Hippel, T., Leggett, S.~K., \& Winget, D.~E.\ 2006, 
\apj, 646, 474  

\bibitem[Krist(2006)]{Krist06}
Krist, J.~E.\ 2006, Spitzer Tiny TIM User's Guide Version 2.0 \\
http://ssc.spitzer.caltech.edu/mips/psf.html 

\bibitem[Kwitter \& Jacoby(1989)]{KJ89} 
Kwitter, K.~B., \& Jacoby, G.~H.\ 1989, \aj, 98, 2159 

\bibitem[Liebert et al.(1989)]{Letal89} 
Liebert, J., et al.\ 1989, \apj, 346, 251

\bibitem[Liebert et al.(2005)]{Letal05} 
Liebert, J., Bergeron, P., \& Holberg, J.~B.\ 2005, VizieR Online Data Catalog, 215, 60047 

\bibitem[McCook \& Sion(1999)]{MS99} 
McCook, G.~P., \& Sion, E.~M.\ 1999, \apjs, 121, 1

\bibitem[Mullally et al.(2007)]{Metal07} 
Mullally, F., et al.\ 2007, \apjs, 171, 206 

\bibitem[Napiwotzki(1999)]{Nap99} 
Napiwotzki, R.\ 1999, \aap, 350, 101 

\bibitem[Napiwotzki(2001)]{Nap01} 
Napiwotzki, R.\ 2001, \aap, 367, 973

\bibitem[Napiwotzki \& Sch\"onberner(1995)]{NS95} 
Napiwotzki, R., \& Sch\"onberner, D.\ 1995, \aap, 301, 545

\bibitem[Nordhaus \& Blackman(2006)]{NB06} 
Nordhaus, J., \& Blackman, E.~G.\ 2006, \mnras, 370, 2004

\bibitem[Phillips(2003)]{Phillips03}
Phillips, J.~P.\ 2003, \mnras, 344, 501

\bibitem[Phillips(2004)]{Ph04} 
Phillips, J.~P.\ 2004, \mnras, 353, 589

\bibitem[Rauch et al.(1999)]{Retal99} 
Rauch, T., K{\"o}ppen, J., Napiwotzki, R., \& Werner, K.\ 1999, \aap, 347, 169

\bibitem[Rauch et al.(1996)]{Retal96} 
Rauch, T., Koeppen, J., \& Werner, K.\ 1996, \aap, 310, 613

\bibitem[Rauch et al.(2009)]{Retal09} 
Rauch, T., Werner, K., Ziegler, M., Koesterke, L., Kruk, J.~W., 
\& Oliveira, C.~M.\ 2009, American Institute of Physics Conference Series, 1135, 171

\bibitem[Reach et al.(2005)]{Reetal05} 
Reach, W.~T., et al.\ 2005, \apjl, 635, L161

\bibitem[Riddle et al.(2011)]{Retal11}
Riddle, A., et al.\ 2011, to be submitted to AJ

\bibitem[Rieke et al.(2004)]{Retal04} 
Rieke, G.~H., et al.\ 2004, \apjs, 154, 25 

\bibitem[Rieke et al.(2005)]{Rietal05} 
Rieke, G.~H., et al.\ 2005, \apj, 620, 1010 

\bibitem[Skrutskie et al.(2006)]{Setal06} 
Skrutskie, M.~F. et al.\ 2006, \aj, 131, 1163

\bibitem[Smith et al.(2007)]{Smith07} 
Smith, J.~D.~T., et al.\ 2007, \pasp, 119, 1133 

\bibitem[Soker(1998)]{Soker98} 
Soker, N.\ 1998, \apj, 496, 833

\bibitem[Su et al.(2006)]{Suetal06} 
Su, K.~Y.~L., et al.\ 2006, \apj, 653, 675

\bibitem[Su et al.(2007)]{Suetal07} 
Su, K.~Y.~L., et al.\ 2007, \apjl, 657, L41

\bibitem[Taam \& Ricker(2010)]{TR10} 
Taam, R.~E., \& Ricker, P.~M.\ 2010, \nar, 54, 65 

\bibitem[Trilling et al.(2008)]{Tetal08} 
Trilling, D.~E., et al.\ 2008, \apj, 674, 1086

\bibitem[Tylenda et al.(1992)]{Tetal92} 
Tylenda, R., Acker, A., Stenholm, B., \& Koeppen, J.\ 1992, \aaps, 95, 337 

\bibitem[von Hippel et al.(2007)]{vHetal07} 
von Hippel, T., Kuchner, M.~J., Kilic, M., Mullally, F., 
\& Reach, W.~T.\ 2007, \apj, 662, 544

\bibitem[Werner et al.(2004)]{Wetal04} 
Werner, M.~W., et al.\ 2004, \apjs, 154, 1 

\bibitem[Werner et al.(1997)]{Wetal97} 
Werner, K., Bagschik, K., Rauch, T., \& Napiwotzki, R.\ 1997, \aap, 327, 721 

\bibitem[Wesemael et al.(1985)]{Wetal85} 
Wesemael, F., Green, R.~F., \& Liebert, J.\ 1985, \apjs, 58, 379

\bibitem[Wyatt(2008)]{wyatt08} 
Wyatt, M.~C.\ 2008, \araa, 46, 339 

\bibitem[Zuckerman \& Becklin(1987)]{ZB87} 
Zuckerman, B., \& Becklin, E.~E.\ 1987, \nat, 330, 138

\bibitem[Zuckerman et al.(2007)]{Zetal07} 
Zuckerman, B., Koester, D., Melis, C., Hansen, B.~M., \& Jura, 
M.\ 2007, \apj, 671, 872 


\end{thebibliography}
\end{document}